\documentclass{aa}
\usepackage[utf8]{inputenc}
\usepackage{newunicodechar}
\newunicodechar{∼}{\textasciitilde{}}
\newunicodechar{α}{\ensuremath{\alpha}}
\newunicodechar{−}{\textminus}
    
\usepackage{graphicx}
\usepackage{subcaption}   
\usepackage{tabularx} 
\usepackage{float} 

\usepackage[dvipsnames]{xcolor}
\usepackage{txfonts}
\usepackage{threeparttable}
\usepackage{xfrac}
\usepackage{dirtytalk}
\usepackage[colorlinks=true,linkcolor=blue,citecolor=blue,urlcolor=blue]{hyperref}
\usepackage{xspace}

\usepackage{makecell}
\usepackage{multirow}
\usepackage{booktabs}
\usepackage{ragged2e}

\newcommand{\kms}{\ensuremath{\,{\rm km}\,{\rm s}^{\rm -1}}\xspace}

\newcommand{\kpc}{\ensuremath{\,{\rm kpc}}\xspace}

\newcommand{\sm}{\ensuremath{\,{\rm M_{\odot}}}\xspace}
\newcommand{\slu}{\ensuremath{\,{\rm L_{\odot}}}\xspace}
\newcommand{\HI}{{\textnormal{H}}{\small \textnormal{I}}\xspace}

\newcommand\cRB[1]{#1}

\begin{document} 

\title{Exploring the stellar streams and satellites around the giant low surface brightness galaxy Malin~1\\
}
\titlerunning{Stellar streams and satellites around Malin~1}

\author{Roy O. E. Bustos-Espinoza}
\date{Submitted to Astronomy & Astrophysics: \today}

   \author{Roy O. E. Bustos-Espinoza
          \inst{1,2} 
          \and
          Matias Bla\~na\inst{3}
          \and
          Gaspar Galaz\inst{1}
          \and
          Marcelo D. Mora \inst{4}
          \and
          Junais\inst{5,6}
          \and
          Mousumi Das\inst{7}
          \and
          Sudhanshu Barway \inst{7}
          \and
          Ankit Kumar \inst{8}
          \and
          Evelyn J. Johnston \inst{9}
          \and
          Thomas Puzia \inst{1}
          }
   \authorrunning{Bustos-Espinoza et al.}
   
   \institute{Instituto de Astrofísica, Pontificia Universidad Católica de Chile,
             Vicuña Mackenna 4860, Macul, Santiago, Chile \\
              \email{robustos@uc.cl}
         \and
             Instituto de Investigaciones F\'isicas, Universidad Mayor de San Andr\'es, La Paz, Estado Plurinacional de Bolivia
        \and 
        Centro Espacial Nacional, Fuerza Aérea de Chile, Av. Pedro Aguirre Cerda 5500, Cerrillos, Santiago, Chile.
        \and  Las Campanas Observatory, Carnegie Observatories, Casilla 601, La Serena, 7820436, Chile
            \and
            Instituto de Astrof\'{i}sica de Canarias, V\'{i}a L\'{a}ctea S/N, E-38205 La Laguna, Spain
            \and
            Departamento de Astrof\'{i}sica, Universidad de La Laguna, E-38206 La Laguna, Spain
            \and
            Indian Institute of Astrophysics, Koramangala, Bangalore 560034, India
            \and
            Instituto de Astrofísica, Departamento de Fisica y Astronomia, Facultad de Ciencias Exactas, Universidad Andres Bello, Fernandez Concha 700, Las Condes, Santiago RM, Chile            
            \and
            Instituto de Estudios Astrofísicos, Facultad de Ingeniería y Ciencias, Universidad Diego Portales, Av. Ejército Libertador 441, Santiago, Chile
             }

   \date{Received \today; accepted \today}

\abstract{
{\textit{Context.} Giant Low Surface Brightness galaxies (gLSBGs), such as Malin~1, host extended stellar and gaseous discs exceeding 100~kpc in radius. Their formation and evolution remain debated, with interactions with satellite galaxies and accretion streams proposed as key contributors. Malin~1 presents multiple nearby and distant satellites. Additionally, it exhibits two giant stellar streams, the largest extending 200 kpc in projection, likely related to past interactions.}

{\textit{Aims.} We investigate the orbital dynamics of Malin~1’s satellites and their possible connections with the observed stellar streams, testing their nature with different formation scenarios.}

{\textit{Methods.} We constructed gravitational potentials for Malin~1 using optical and \HI{} rotation curve data, incorporating stellar, gaseous, and dark matter (DM) components. We explored a wide orbital parameter space to determine whether the candidate progenitors of the stellar streams could have originated from past interactions, testing both Navarro-Frenk-White (NFW) and the pseudo-isothermal (ISO) halo profiles.}

{\textit{Results.} Among several explored scenarios, some produced bound orbital solutions. The ISO halo model, with mass of $M_{\rm Virial} \approx 2.6 \times 10^{12}\sm$, favours bound satellite orbits more than the NFW model, that has a lower mass of $M_{\rm Virial} \approx 1.4 \times 10^{12}\sm$. 
These orbital models show that the giant stellar streams could be substructures of some satellites galaxies located along their leading and trailing orbital trajectories.
Furthermore, our analysis indicates that the most distant Malin~1 satellite (eM1) could have reached its orbital pericentre $\sim1.6$ Gyr ago, 
while the nearest companions could have interacted as early as $\sim100$ Myr ago.
At the same time, one of the closest companions would still be under a strong interaction.
Another close companion displays both leading and trailing arms in a radial orbit, although it also shows a polar orbit. Furthermore, we identify some unbound orbital solutions that could link some satellites with streams.}

{\textit{Conclusions.} 
The observed alignment of satellites and streams indicates that past interactions likely shaped aspects of Malin~1’s morphology. Our orbital modelling constrains possible progenitors of the observed stellar streams and their orbital histories, providing new insights into the dynamical evolution of gLSBGs. Our findings are consistent with results reported very recently by other studies using Malin~1 kinematic data.}
}


\maketitle
\nolinenumbers
%

\section{Introduction}
\label{sec:intro}


\citet{Disney1976} suggested the possibility that there are galaxies with a surface brightness (SB) lower than the empirical limit proposed by \citet{Freeman1970ONGALAXIES}, which defined the characteristic central surface brightness of disc galaxies as being around $21.65$ $B$ mag arcsec$^{-2}$. This threshold was based on observations of high-surface-brightness spirals and was initially thought to represent a universal feature of disc galaxies.
Since then, the discovery rate of low surface brightness galaxies (LSBGs) has been boosted by advances in observational techniques and deeper sky surveys, leading to a growing recognition of their importance in galaxy formation and evolution \citep{Du2020}.

In general terms, LSBGs exhibit a central surface brightness fainter than $23.5$ $B$ mag arcsec$^{-2}$ \citep{Impey1997}. These galaxies display poor star formation rates, a low metallicity, and extended \HI{} gas discs, characteristics shared with giant low surface brightness galaxies (gLSBGs), which appear as massive systems typically existing in isolation and have sizes larger than that of the Milky Way up to the 25th isophote \citep{Das2013}. 

Malin~1, one of the most iconic gLSBGs due to its immense size and faintness, was discovered serendipitously by \citet{Bothun1987}. This particular gLSBG has a bar structure and shows a LINER nucleus \citep{Barth2007}. Evidence indicates that the bar resulted from several star formation events, and along with the discovery of compact sources,
the same evidence points to a possible double nucleus system \citep{Johnston2024}. 
\citet{Ogle2016, Ogle2019} reported the discovery of a large sample of the most optically luminous, largest, and most massive spiral galaxies in the Universe, called super spirals, that could be analogous to Malin~1; however, Malin~1 is still the largest stellar disc known \citep{Ogle2016, Ogle2019}, but it has a much lower star formation rate of $\sim 1.2\,M_{\odot}\,\mathrm{yr}^{-1}$ \cRB{than those observed in typical super spirals} \citep{Lelli2010, Junais2024MUSEAttenuation}.

Malin~1 also presents HII regions, a radial decrease in the surface density of the star formation rate, reduced metal content, and flattening in the outer disc \citep{Junais2024MUSEAttenuation}. These characteristics provide clues about the development of the disc, with the most likely scenario involving the accretion of pre-enriched gas from a previous merger \citep{Junais2024MUSEAttenuation, Zhu2018}. Furthermore, the galaxy is characterised by a low dust content and a minimal presence of molecular gas \citep{Gerritsen1999, Galaz2022, Junais2024MUSEAttenuation, Galaz2024}.

Moreover, an optical map has revealed a cavity in the southern part \cRB{of Malin 1}, between the arms \citep{Galaz2015}. This cavity could be due to feedback from star formation, as the \HI{} map shows a nearby asymmetric substructure \citep{Lelli2010}, or the action of supernova-driven winds expelling gas, similar to phenomena observed in other galaxies such as the Hoag galaxy \citep{Bannikova2018}, the Fireworks galaxy \citep{Blair2019}, NGC~1300  \citep{Maeda2022}, or the gLSBG UGC 1382 \citep{Saburova2022aems.conf..395S}. 
A northern warp in the \HI{} disc shows axis asymmetry, possibly due to the accretion of a paired system of two gas-rich satellite galaxies. This \cRB{could} be evidence of ongoing interaction \citep{Saha2021}. 

\begin{figure*}[t]
\sidecaption
  \includegraphics[width=11.5cm]{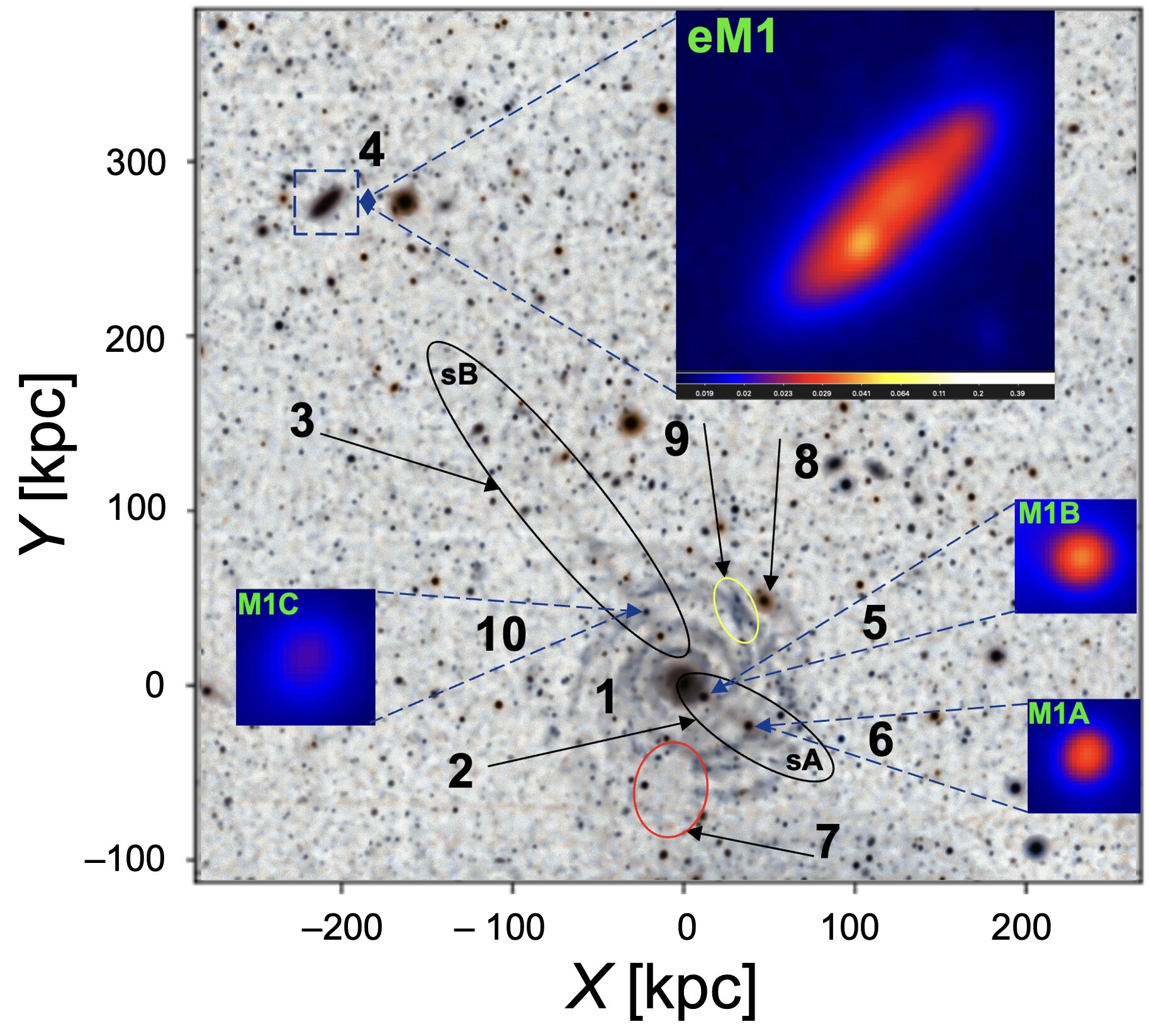}
  \caption{Malin~1 system $g$ and $r$ band image from the Magellan/Clay telescope \citep{Galaz2015}. 
  The equatorial coordinates \cRB{have been} converted into $X$, $Y$ in \cRB{kiloparsec} units, adopting an angular scale of $1.557$ kpc per arcsec$^{-1}$, \cRB{and} the origin \cRB{is} centred at Malin~1. 
  The \cRB{colour-map} scale \cRB{for the sub-panels uses} arbitrary units \cRB{and is} shown in  panel 4. Labels mark \cRB{the following}: {1.} Malin~1.
  {2.} Proposed stream A (sA), \cRB{possibly} linking Malin~1A, Malin~1B, and Malin~1. 
  {3.} Proposed stream B (sB), aligned in projection with exo-Malin~1, Malin~1C, and Malin~1. 
  {4.} The satellite galaxy SDSS J123708.91+142253.2 \citep{Reshetnikov2010}, referred \cRB{to} here as exo-Malin~1 (eM1) at coordinates $X,Y=(-211,277)$ kpc, separated $\sim 350 \;$ kpc from Malin~1. 
  Note \cRB{the} offset between the brightness peak and the geometrical centre (zoom-in). 
  {5.} Malin~1B (M1B)
  {6.} Malin~1A. \cRB{Both, Malin 1B and 1A} are compact ellipticals, and interacting galaxies \citep{Saburova2022aems.conf..395S}.
  {7.} Cavity, \cRB{a low stellar density region displaying an apparent gap between the spiral arms} (red ellipse).
  {8.} A distant background galaxy ($z$=0.3639). 
  {9.} \cRB{The yellow} ellipse indicates clumps within the spiral arms; their structure suggests regions of star formation or newly formed star clusters, based on VLT-MUSE observations \citep{Junais2024MUSEAttenuation}. 
  {10.} A UDG candidate \citep{Ji2021}, with VLT-MUSE data \citep{Junais2024MUSEAttenuation} showing a young star formation region with redshift $z=0.08292$. It is $\sim 47$ kpc from Malin~1 centre, and \cRB{we refer to} it as Malin~1C ({M1C)}.
  }
  \label{Fig:Malin1EnvironmentIn}
\end{figure*}

Studies \cRB{have revealed} that Malin~1 interacts with at least two companion galaxies, identified as Malin~1B and the SDSS J123708.91+142253.2 \citep{Reshetnikov2010, Galaz2015} - the latter is also referred to as exo-Malin~1, a designation initially introduced by \citet{Bustos2024DDA....5540520B, Bustos2024IAUGA..32P2541B}, or eM1 for brevity. \cRB{Both galaxies} are associated satellite galaxies. In addition, Malin~1A is part of the Malin~1 system and, together with Malin~1B, is classified as a compact elliptical \citep{Saburova2022aems.conf..395S}. A recent study by \citet{Junais2024MUSEAttenuation} identified several ${\rm H\alpha}$ regions, including Malin~1C, which is part of the Malin~1 galaxy and may be an ultra-diffuse galaxy (UDG) that interacts with the core galaxy \citep{Ji2021}.
Malin~1C has possibly been stripped of gas and outlying stars by tidal interactions \citep{Kazantzidis2004ASPC..327..155K}. Malin~1B and exo-Malin~1 are situated at distances of $14$~kpc and $350$~kpc \cRB{towards} the south-east and north-west, respectively \citep{Reshetnikov2010}. These objects, along with others, are illustrated in detail in Fig. \ref{Fig:Malin1EnvironmentIn}. 

According to \citet{Saha2021}, exo-Malin~1 may play a role in star formation within the central bar, while the accretion of Malin~1B could bolster star formation in the arms and bulge \citep{Saha2021}. As noted by \citet{Saburova2023} and \citet{Penarrubia2006}, extended discs in gLSBGs could develop through minor mergers involving gas-rich satellites, \cRB{and this scenario is} possibly applicable to exo-Malin~1, Malin~1B, Malin~1A, and Malin~1C. 

Tidal stellar streams exemplify the continuous merging and accretion of galaxies, a key process in the formation of structures within the Universe \citep{Niederste2012}. Several well-studied examples \cRB{have illustrated} the remnants of disrupted dwarf galaxies and their tidal debris. The Sagittarius stream, which originates in the dwarf galaxy of Sagittarius, demonstrates that progenitors can lie several to tens of degrees away from their associated streams \citep{Niederste2012, Bonaca2021}. Similarly, a vast stellar stream extending more than 500 kpc has been identified on the outskirts of the Coma cluster. \cRB{The stream} likely formed from the destruction of a dwarf galaxy\cRB{, and it provides} key insights into the gravitational potential of the cluster \citep{Roman2023}. The spatial relationship between a satellite and its debris depends on its orbital phase\cRB{. Near} the apocentre, the material tends to accumulate into overdensities, while near the pericentre, the tidal stream stretches along the orbit and compresses perpendicular to its motion due to its acceleration and host tidal field \citep[e.g.][]{Niederste2012,Smith2013,Blana2015}.

Analysis of the luminosity of the stream progenitors suggests that they were absorbed during relatively early formation periods \citep{Vera-Casanova2022}. Moreover, recent studies of the giant spiral galaxy \ion{Malin}{2} \cRB{have} revealed faint tidal structures surrounding the main host \citep{Junais2025}, suggesting that tidal interactions and mergers may be common processes in ultra-low surface brightness systems.

Various hypotheses \cRB{have been developed} here regarding the origin of gLSBG\cRB{s}. According to \citet{Noguchi2001}, gLSBGs evolve\cRB{d} from standard spiral galaxies through dynamic changes over time. Their formation is influenced by environmental factors, including large galaxies in voids that feature conventional bulges coupled with extensive LSB discs \citep{Hoffman1992}. Within the hierarchical formation paradigm, mergers between a central galaxy and its satellite companions may explain the considerable size observed \citep{Penarrubia2006}. Although the origins and properties remain partially unknown, some cosmological simulations point to these alignment-driven mergers as the key to such large-scale development \citep{Zhu2023}. Moreover, \citet{Zhu2018} identified that a significant fraction of the cold gas at redshift zero originated from the cooling of the hot halo gas, a process triggered by the merger of a pair of interacting galaxies.

The Lambda cold dark matter ($\Lambda$CDM) cosmology theory suggests that the disruption of satellite galaxies is a frequent process in the lifetime of massive galaxies \citep{Martinez-Delgado2023, Miro2024}. The extended size of the \cRB{gas and stellar} disc might result from the impact of two or more galaxies merging. This formation mechanism aligns with the existing theories of formation of gLSBGs \citep{Zhu2018}. As \citet{Roman2023} pointed out, such giant faint stellar streams may exist in galaxy clusters and groups according to $\Lambda$CDM. So the presence of streams in the Malin~1 system could be evidence that not all gLSBGs are isolated, in contrast to \citet{Das2013}, who argued that gLSBGs are usually isolated, often near the edges of voids, and that their lack of interactions contributes to their slow evolution. Research involving cosmological simulations \citep{Zhu2018} illustrates that smaller galaxies, often termed intruders, can interact with larger galaxies, leading to the formation of a gLSBG that shares notable similarities with Malin~1. Similarly, \citet{Mapelli2008MNRAS.383.1223M} proposed that stimulated accretion might explain the exceptionally large disc size of Malin~1. Additionally, \citet{Zhu2018} indicated that a collision involving three galaxies could produce a gas-rich \cRB{gLSBG}.

This paper investigates Malin~1 and its surroundings. We analysed possible orbital configurations of its satellite galaxies, including both leading and trailing segments, to comprehend their impact on its structure, kinematics, and evolutionary progression.

The structure of this article is as follows.
In \cRB{Sect.} \ref{sec:intro:malin1in} \cRB{we} describe the observed substructures.
Section \ref{sec:method} details the modelling approach for the Malin~1 system. 
Section \ref{sec:method:obs} provides the observational data and associated constraints. 
\cRB{In Sect.} \ref{sec:Malin1GravitationalPotential} \cRB{we discuss} the gravitational potential of Malin~1, and Section \ref{sec:orbitalmodelling} outlines scenarios to explore progenitor candidates of the stellar streams, including initial setups. \cRB{In Sect.} \ref{sec:res} \cRB{we} evaluate the results, and \cRB{in} Section \ref{sec:discuss} \cRB{we discuss and interpret} our findings. The conclusion is found in Section \ref{sec:conclu}. \cRB{Finally}, two appendices \cRB{provide a summary of} the optimal parameters and describe how prior knowledge is integrated into our model.

For our calculations, we used a flat $\Lambda$CDM cosmology with H$_0$ set at 70 km s$^{-1}$ Mpc$^{-1}$, $\Omega_{\rm M} = 0.27$, and $\Omega_{\Lambda} = 0.73$. This \cRB{resulted} in a projected plate scale of 1.557 kpc arcsec$^{-1}$ and a luminosity distance to Malin~1 of 377 Mpc.

\section{Observed substructures in the Malin~1 system} 
\label{sec:intro:malin1in}

This section presents a summary of the features identified in the optical image of Malin~1 \citep{Galaz2015}. The environment of Malin~1, observed in the $g$ and $r$ bands with the $6.5$ m \textit{Magellan/Clay} telescope, is shown in Fig.~\ref{Fig:Malin1EnvironmentIn}, and the listed substructures are indicated by arrows that follow the same numbering. The essential characteristics are listed in Table \ref{table:M1PhysicalProperties}. Among the relevant properties of the Malin~1 system, the following are noteworthy:

\begin{table}[ht]
\caption{Properties of the Malin 1 galaxy.}
\label{table:M1PhysicalProperties}
\centering
\begin{tabular}{ll}
\hline\hline
Property & Value \\
\hline
Morphology & SB0/a \tablefootmark{a} \\
$D_L$ [Mpc] & $377 \pm 8$ \tablefootmark{b} \\
$M_V$ [mag] & $-22.9 \pm 0.4$ \tablefootmark{c} \\
$\mu_{0,B}$ [mag arcsec$^{-2}$] & $\sim 28.0$ \tablefootmark{d} \\
$\mu_{0,V}$ [mag arcsec$^{-2}$] & $\sim 25.5$ \tablefootmark{e} \\
$M_{\ion{H}{i}}$ [$10^{10}$ M$_{\sun}$] & $4 - 7$ \tablefootmark{f} \\
AGN & LINER \tablefootmark{g} \\
$\alpha$ (J2000) & $12^{\mathrm{h}} 36^{\mathrm{m}} 59.35^{\mathrm{s}}$ \tablefootmark{h} \\
$\delta$ (J2000) & $+14^\circ 19' 49.16''$ \tablefootmark{h} \\
$z$ & $0.0827002$ \tablefootmark{h} \\
$V_{\odot}^{\star}$ [km s$^{-1}$] & $24775 \pm 10$ \tablefootmark{h} \\
$D_{\text{opt}}$ [kpc] & $\sim 220$ \tablefootmark{i} \\
$i$ & $38^\circ \pm 3^\circ$ \tablefootmark{j} \\
\hline
\end{tabular}
\tablefoot{
\tablefoottext{a}{Barred lenticular galaxy with faint spiral arm features \citep{Lelli2010}.}
\tablefoottext{b}{\citet{Lelli2010}.}
\tablefoottext{c}{\citet{Pickering1997}.}
\tablefoottext{d}{\citet{Galaz2015}.}
\tablefoottext{e}{\citet{Impey1997}.}
\tablefoottext{f}{\citet{Pickering1997, Lelli2010}.}
\tablefoottext{g}{\citet{Barth2007}.}
\tablefoottext{h}{Equatorial coordinates, redshift, and velocity from \citet{Reshetnikov2010}.}
\tablefoottext{i}{Optical and \ion{H}{i} diameter, scaled to the Milky Way \citep{Moore2006}.}
\tablefoottext{j}{\citet{Lelli2010}.}
}
\end{table}

\begin{table*}
\centering
\caption{Line of sight velocity $V_{\mathrm{los}}$, and related properties of \textit{Malin~1} and associated galaxies}
\label{table:RedshiftsA}
\begin{tabular}{llllll}
\hline\hline
Galaxy \tablefootmark{a} & $V_{\mathrm{los}}$ [km s$^{-1}$] \tablefootmark{b} & Petro. radius [kpc] \tablefootmark{g} & Luminosity [$L_{\odot}$] \tablefootmark{h} & $\gamma_{\mathrm{band}}$ [$M_{\odot}/L_{\odot}$] \tablefootmark{n} & $M_{\star}$ [$M_{\odot}$] \tablefootmark{q} \\
\hline
M1  & $0 \pm 1$ \tablefootmark{c}    & $5.6 \pm 0.2$ & $(1.07 \pm 0.06) \times 10^{11}$ \tablefootmark{i} & $3.3 \pm 0.4$ \tablefootmark{i} & $(1.4 \pm 0.3) \times 10^{11}$ \tablefootmark{r} \\
eM1 & $132 \pm 37$ \tablefootmark{d} & $5.2 \pm 0.4$ & $(5.6 \pm 0.3) \times 10^{9}$ \tablefootmark{d}    & $0.9 \pm 0.2$ \tablefootmark{p} & $(5.0 \pm 0.9) \times 10^{9}$ \\
M1B & $65 \pm 22$ \tablefootmark{d}  & $2 \pm 2$     & $(2.6 \pm 0.7) \times 10^{9}$ \tablefootmark{d}    & $3.4 \pm 0.6$ \tablefootmark{o} & $(8.8 \pm 1.7) \times 10^{9}$ \\
M1C & $54 \pm 27$ \tablefootmark{e}  & $1.5 \pm 0.5$ & $(1.3 \pm 0.2) \times 10^{8}$ \tablefootmark{l}    & $0.9 \pm 0.2$ \tablefootmark{p} & $(1.2 \pm 0.3) \times 10^{8}$ \\
M1A & $54 \pm 27$ \tablefootmark{f}  & $2 \pm 2$     & $(1.12 \pm 0.15) \times 10^{9}$ \tablefootmark{m}  & $3.4 \pm 0.6$ \tablefootmark{o} & $(3.8 \pm 0.7) \times 10^{9}$ \\
\hline
\end{tabular}
\tablefoot{
\tablefoottext{a}{Galaxy abbreviations defined in Table~\ref{table:ScenariosUnified} and Fig.~\ref{Fig:Malin1EnvironmentIn}.}
\tablefoottext{b}{$V_{\mathrm{los}}$ velocity of the satellite with respect to the Malin 1 system.}
\tablefoottext{c}{M1 centre with dispersion from the projected angular scale \citep{Galaz2015}.}
\tablefoottext{d}{Reported by \citet{Reshetnikov2010}.}
\tablefoottext{e}{\citet{Junais2024MUSEAttenuation} and private communication.}
\tablefoottext{f}{Our estimation based on both M1A and M1C having approximately the same distance from the M1 centre.}
\tablefoottext{g}{Radius where surface brightness is 20\% of the average within it \citep{Petrosian1976, Psychogyios2016}; data from \citet{Sloan2024}.}
\tablefoottext{h}{Luminosity in Solar units.}
\tablefoottext{i}{This work: averaging isothermal and NFW profiles.}
\tablefoottext{l}{UDG in $r$-band \citep{Sloan2024}.}
\tablefoottext{m}{cE in $r$-band \citep{Sloan2024}.}
\tablefoottext{n}{Mass-to-light ratio in solar units \citep{Bell2001}.}
\tablefoottext{o}{$\gamma$ value for cE \citep{Bell2001}.}
\tablefoottext{p}{$\gamma$ value for LSB or UDG \citep{Bell2001}.}
\tablefoottext{q}{Stellar mass.}
\tablefoottext{r}{This work: stellar mass estimation within $r < 100$~kpc, averaging models with ISO and NFW profiles.}
}
\end{table*}

\begin{itemize}
\item[-] {1. Malin~1:} The gLSBG comprises a dual structure: a central bulge-bar and HSB disc resembling a Sa/SB galaxy with a radius under 10 kpc, and an extended, gas-rich outer LSB region with an optical diameter of 240 kpc \citep{Barth2007, Moore2006}. This last component features the most extensive known stellar disc, reaching a surface brightness of approximately 28.0 $B$ mag arcsec$^{-2}$ \citep{Galaz2015, Boissier2016}. The galaxy also possesses a large \HI{} gas reservoir with a mass $4-7 \times 10^{10} M_{\odot}$ \citep{Pickering1997, Lelli2010}. It also has a high specific angular momentum \citep{DiTeodoro2023, Boissier2016, Salinas2021TheGalaxies}, and an extended flat rotation curve \citep{Lelli2010, Pickering1997}, all of which suggest the presence of a vast dark matter (DM) halo \citep{Mo2010}. 

\item[-] {2. Stream A (sA):} As noted by \citet{Galaz2015}, \cRB{it is} a possible stellar stream that \cRB{may be} associated with two galaxies: Malin~1B, positioned at $\sim 14$ kpc from the centre of Malin~1 (marked as point 5), and \cRB{(more likely)} Malin~1A, which sits in the middle of the stream at $\sim 45$ kpc from Malin~1 centre (marked as point 6), {both in the south-west direction}. 

\item[-] {3. Stream B (sB):} The largest ellipse shown in Fig. \ref{Fig:Malin1EnvironmentIn} outlines a structure found in \citet{Galaz2015} {with an elongation that suggests a stellar stream, a jet, or a diffuse region \citep{Boissier2016}. Here, we call this stellar stream candidate Stream B.
sB is significantly extended, reaching up to $200$~kpc from the centre of Malin~1 in the direction of the satellite exo-Malin~1.
If this extended feature corresponds to a stellar stream, then it could have formed from material—either stellar or gaseous—tidally stripped from a satellite galaxy.
Here, we explore progenitors that could have had past encounters with Malin~1, removing material from the satellite, producing the stream-like structure.}

\item[-] {4. exo-Malin~1 (eM1):} \cRB{This galaxy has a} total luminosity of $9.47 \times 10^9 L_{\odot}$. While it shows an elongated morphology, it \cRB{is} unclear if this corresponds to an early-type galaxy or a disc structure. 
{It shows evidence of emission lines of H$\alpha$, H$\beta$, and OII, so it is likely a star-forming disc galaxy \citep{Sloan2024, Lintott2011}.}
Its redshift of $z=0.083082$ and projected distance of $350\kpc$ makes it a Malin~1 satellite candidate, that may have interacted in the past \citep{Reshetnikov2010}. This interaction could explain the formation of the extended low surface brightness (LSB) disc or serve as a catalyst for the emergence of the suspected stream B \citep{Galaz2015}. Additionally, the zoom-in image of eM1 (see {point 4} in Fig. \ref{Fig:Malin1EnvironmentIn}), shows that the peak of the surface brightness is shifted from the geometrical centre. 
This off-centre, found in this study, could indicate that exo-Malin~1 has experienced past interactions that could involve Malin~1. Further discussion on this can be found in the Section \ref{sec:discuss}. Moreover, it is very intriguing that eM1 appears to align on the sky with the stream~B along an almost straight line towards the centre of Malin~1.

\item[-] {5. Malin~1B (M1B):} \citet{Reshetnikov2010} recognised it as a galaxy interacting with Malin~1 (point 5 in Fig. \ref{Fig:Malin1EnvironmentIn}). Furthermore, \citet{Saha2021} identified its interaction specifically with the central area of Malin~1, implying it might be responsible for initiating recent star formation in the galaxy \citep{Saha2021}. The possibility that stream A is concurrently connected with Malin~1B \citep{Reshetnikov2010} and/or Malin~1A \citep{Saburova2022aems.conf..395S} supports the hypothesis of a past merger. Malin~1A and Malin~1B, both compact ellipticals \citep{Saburova2022aems.conf..395S}, could have experienced tidal stripping, potentially explaining their compact nature. If further stripping occurs, they might evolve into objects similar to ultra-compact dwarfs (UCDs), as suggested by \citet{Bekki2001ApJ...552L.105B}. 
Moreover, such tidal shaping of satellite galaxies has been shown to operate on mass scales of magnitudes of orders of difference \citep{Blana2025}.
Meanwhile, Malin~1C's classification remains unclear: \cRB{It} could be either a UDG \citep{Ji2021} or a young star-forming region \citep{Junais2024MUSEAttenuation}, possibly linked to past interactions.

\item[-] {6. Malin~1A (M1A):} \cRB{This object was} identified as another compact elliptical, \cRB{which are} typically located close to larger galaxies and prone to tidal interactions and stripping \citep{Saburova2022aems.conf..395S, Johnston2024}. M1A has undergone significant tidal encounters with Malin~1, leading to the depletion of nearly all their gas \citep{Kim2020ApJ...903...65K}. 

\item[-] {7. Cavity:} A cavity in the southern region of the observed LSB disc indicates a reduced stellar density, displaying an apparent gap between spirals. Furthermore, irregularities in the \HI{} gas distribution observed in the \HI{} data are also noted \citep{Lelli2010}. Its origin is unknown: it could be the result from a local gas depletion, dust extinction, or satellite interactions.  

\item[-] {8. Galaxy SDSS J123657.22+142021.1:} \cRB{A} background galaxy at a high redshift ($z=0.364$) beyond Malin~1, and a broadline galaxy \citep{Abazajian2009}.

\item[-] {9. HII regions:} Identified using VLT-MUSE, such as the one highlighted by a yellow ellipse, indicate recent star formation activity \citep{Junais2024MUSEAttenuation}.

\item[-] {10. Malin~1C (M1C):} {\cRB{R}eported as an star forming (${\rm H\alpha}$) blob candidate} using VLT-MUSE observations \citep{Junais2024MUSEAttenuation}, with signatures of ${\rm H\alpha}$ emission and  at a redshift similar to Malin~1 ($z=0.08292$).
It is positioned at a projected distance of $46.75$ kpc from Malin~1, forming part of the Malin~1 surroundings. 
Fig.~\ref{Fig:Malin1EnvironmentIn} shows a \cRB{colour}-enhanced image of M1C.
MUSE ${\rm H\alpha}$ emission line has a relative mean velocity of $54$ km/s from Malin~1 systemic velocity, {being likely bound to Malin~1 and with a velocity dispersion higher than 90 km/s} \citet[][private communication]{Junais2024MUSEAttenuation}. 
{This high velocity dispersion could also mean that the system corresponds to a dwarf galaxy.
Moreover, the nature of ${\rm H\alpha}$ blobs are still debated, with tidal remnants, stripped clouds, or \cRB{UDGs} as candidates \citet{Ji2021}.}
Moreover, M1C is appears aligned {with the stream B, which would be a natural consequence if this is the core of the stream that is being tidally disrupted. Current orbital configurations significantly influence tidal stream arrangements, impacting both the leading and trailing arms, and possibly affecting the remnant bound core. This effect is supported by various simulation studies \citep[see Fig. 1 in][]{Niederste2012}}.

\end{itemize}

\section{Modelling the Malin~1 system}
\label{sec:method}
Here we describe in detail how we \cRB{modelled} the gravitational potential of Malin~1 and the possible orbits of its satellites and streams.
The gravitational potential and orbital models are constrained by observational data, such as the surface brightness profile, rotation curve, and line-of-sight velocities.

\begin{table*}
\centering
\caption{Definition of parameters used in the orbital modelling for each scenario.}
\label{tab:parameters}
\begin{tabular}{l l p{10cm}}
\hline\hline
Parameter & Symbol & Description \\
\hline
Spatial coordinate (los) & $Z$ & Line-of-sight distance of the progenitor candidate relative to the mid-plane of Malin~1 (kpc). A positive (negative) value places the object in front of (behind) the disc along the observer’s line of sight. \\
Tangential velocity & $V_{\mathrm{tg}}$ & Magnitude of the velocity component perpendicular to the line of sight (km\,s$^{-1}$). Together with the observed los velocity, it defines the full 3D velocity vector of the orbit. \\
Position angle & PA & Angle measured at the orbit initial condition, indicating the direction of the tangential velocity vector in the plane of the sky, measured counterclockwise from north through east (degrees). \\
\hline
Pericentre & $R_{\mathrm{peri}}$ & Minimum galactocentric distance to Malin~1 ($r$) reached by the orbit (kpc). \\
Apocentre & $R_{\mathrm{apo}}$ & Maximum distance $r$ to Malin~1 reached by the orbit (kpc). \\
Eccentricity & $\epsilon$ & Orbital shape, defined as $\epsilon = (R_{\mathrm{apo}} - R_{\mathrm{peri}}) / (R_{\mathrm{apo}} + R_{\mathrm{peri}})$. Circular orbits have $\epsilon=0$ and highly elongated orbits have $\epsilon \to 1$.\\
Posterior probability & $\log(p)$ & Logarithm of the posterior probability returned by the \textsc{mcmc} sampling. Larger values correspond to orbital solutions that are statistically more significant. \\
\hline
\end{tabular}
\tablefoot{The scenarios mentioned are described in detail in Table~\ref{table:ScenariosUnified}.}
\end{table*}

\subsection{Observational data and constraints}
\label{sec:method:obs}

For the photometric modelling, we considered different surface brightness (SB) band data observations of Malin~1: $HST/F814W$ for the central area, within $<15$ kpc \citep{Barth2007}, and the $r$ and $g$ bands for the extended disc, reaching $100 - 120$ kpc in radius \citep{Moore2006, Reshetnikov2010, Saha2021}. In the analysis, we also used the SB profile of the $\textit{R}$-band as explained in \citet{Lelli2010}. Analysis was conducted on the \HI{} rotation curve data \citep{Pickering1997, Lelli2010}, taking into account the stellar, gaseous, and DM mass components. {Furthermore, the inner part of the \cRB{rotation curve}, dominated by the bulge/core and HSB disc, would be constrained by incorporating existing kinematic data of the central region, such as long-slit spectroscopy \citep{Junais2020}, MUSE observations \citep{Johnston2024, Junais2024MUSEAttenuation, Saburova2022aems.conf..395S}}. 

In addition, the sky coordinates of the stellar streams, measured with respect to the centre of Malin~1 as the inertial reference frame, were derived from the deep photometric maps of \citet{Galaz2015}, who report a $\sim 200$ kpc stream that extended \cRB{towards} exo-Malin~1. We also carried out a visual inspection of these maps, guided by surface brightness enhancements relative to the background, to select representative coordinates $x$-$y$ points along both streams, which were then used as observational constraints in our orbital modelling. The $x$-axis aligns with the right ascension ($\alpha$) and increases leftward on the sky plane. The $y$-axis aligns with the declination ($\delta$) and increases northward (upward) in the plane of the sky. The statistical accuracy of each coordinate $x$ and $y$ was enhanced by using the projected angular scale and the median seeing of $0.8$ arcsec \citep{Galaz2015}, given a confidence level of $\pm 1.25$ kpc. For the distance $r$ measured from the centre of Malin~1, error propagation was used (see Fig. \ref{Malin1_AandBstreamPoints} and Table \ref{Table:streamPoints}). 

Moreover, the line of sight velocity of the satellite, with respect to the Malin~1 system, denoted $V_{\text{los}}$ and cited from various sources \citep{Dey2019, Cook2023, Reshetnikov2010, Saburova2022aems.conf..395S, Galaz2015, Junais2024MUSEAttenuation, Johnston2024}, along with other likelihood-associated satellite galaxies, is listed in Table \ref{table:RedshiftsA}.

\subsection{Malin~1 gravitational potential}
\label{sec:Malin1GravitationalPotential}

To investigate the dynamics of the Malin~1 satellites and streams, we need to model the gravitational potential to calculate the orbits. 
For this, we proceed with the modelling process in two steps.

{First, we construct a model of the surface brightness profile of Malin~1 using the photometric data described in Section \ref{sec:method:obs}. This model represents the main Malin~1 substructures described in Section \ref{sec:intro:malin1in}: the bulge/bar and the HSB and LSB discs. We adopted photometric models that offer analytical solutions for their potentials, which allow for quick orbital calculations and a large parameter space exploration. Therefore, we used the Miyamoto-Nagai (MN) model proposed by \citet{Miyamoto1975}, where its potential relates to the density through the Poisson equation and is defined as:
\begin{equation}
    \Phi_{MN}(R,z) = -\frac{GM}{\sqrt{R^2 + \left(a + \sqrt{Z^2 + b^2}\right)^2}},
    \label{eq:MNpotential}
\end{equation}
where $a$ and $b$ are the radial and vertical scale lengths, $M$ is the mass, 
$R=\sqrt{X^2+Y^2}$ is the cylindrical radius, $X, Y, Z$, and $r=\sqrt{R^2+Z^2}$ is the spherical radius. 
We note that when $a = 0$, the potential $\Phi_{MN}$ simplifies to the Plummer profile \citep{Plummer1911}.
For photometric modelling, we fit the total luminosity of each component $L_x$. The luminosity can be converted to stellar mass $M_{\star} = \gamma_{\rm band} L_{\rm band}$, where $\gamma_{x}$ is the stellar mass-to-light ratio in a band. In our analysis, we used a Plummer profile to model the bulge/bar of Malin~1 and two MN models for the components of the HSB and LSB disc. We also \cRB{tested} replacing the spherical bulge model with a triaxial bar model \citet{Long1992} using \cRB{the} Plummer luminosity and scale lengths from \citet{Saha2021}, finding similar main orbital properties. Given that the MN profile becomes a Kuzmin profile if a face-on disc is integrated along the \cRB{line-of-sight (los)} \citep{Binney2008}, we used this simplified profile to fit the surface brightness profiles.

In the second step, we fitted circular velocity ($v_{\rm c}$) models to the observed gas kinematics to obtain the mass-related parameters.
Here we fit the mass-to-light ratio $\gamma$ under the maximal disc assumption, which means that the stellar components contribute the maximum possible fraction of the observed rotation curve before the DM halo becomes dominant \citep{Binney2008}. 
For the representation of the dark halo, we implemented two sets of models. 
One with the Navarro-Frenk-White (NFW) profile \citep{Navarro1996}, which aligns with the $\Lambda$CDM model, and another with a pseudo-isothermal (ISO) profile \citep{DeBlok2002}.
Both profiles are widely used to describe the density distribution of DM halos, as has been applied to Malin~1 \citep[e.g.][]{Lelli2010}. NFW assumes that \cRB{DM} halos have a characteristic 'cuspy' central density and extend over large distances. The ISO profile, on the other hand, features a flat central core, which implies a constant density at the centre \citep{Mo2010}. These are defined as

\begin{equation}
\Phi_{NFW}(r) = - \frac{GM_h \ln(1 + r/r_s)}{r} \left[ \ln(1 + c) - \frac{c}{1 + c} \right]^{-1},
\label{eq:NFW}
\end{equation}

\begin{equation}
    \Phi_{Iso}(r) = -4\pi G \rho_0 r_c^2 \left[\frac{1}{2} \ln\left(1 + \frac{r^2}{r_c^2}\right) + \frac{r_c}{r} \tan^{-1}\left(\frac{r}{r_c}\right)\right].
\label{eq:IsoThermal}    
\end{equation}

The NFW model defines the concentration parameter $c = r_{200}/r_s$, the virial radius $r_{200}$, where the mean density becomes 200 times the critical density of the Universe, and the scale radius $r_s$. 
On the other hand, in the ISO profile \citep{DeBlok2002}, the parameter $r_c$ defines the radius of the core and $\rho_0$ indicates the central density of the DM halo.

In addition, Malin~1 has a very extended \HI{} disc component \citep{Bothun1987, Pickering1997}. 
To include this in the gravitational model, we again used a Miyamoto-Nagai model with a mass of $M_{\HI{}} = (4-7) \times 10^{10} M_{\odot}$ \citep{Lelli2010}, a scale length of $a_{\HI{}}=a_{2}$, where we assumed that it has a distribution similar to that of the LSB component and a height of $z_{\rm gas}=0.3\kpc$.
Furthermore, factors 1.1 for molecular $H_2$ and 1.4 for helium gas are included \citep{Lelli2010}. Together, these components represent the baryonic mass.

\subsection{{Orbital modelling: \cRB{S}etup, parameters, and methodology}}
\label{sec:orbitalmodelling}

Literature studies on the disruption of satellite galaxies and the formation of stellar streams with N-body simulations typically find that the remaining satellite core has a leading and a trailing component \citep[e.g.][and references therein]{Fellhauer2006,Smith2013}. 
Moreover, satellites and their streams can have different morphologies that depend on the potential of the host and, more importantly, on the eccentricity of the orbit. Very circular orbits can produce ring-like streams or polar ring streams, as observed in the Sombrero galaxy \citep{Martinez-Delgado2021}. Radial orbits can produce a more diverse morphology depending on the location where the satellite and the streams are observed in their common orbit.
For example, a satellite can be preferentially closer to one of its streams, with the trailing stream being closer to the apocentre, and the leading arm could be tidally compressed while entering the host, with the bound core somewhere in the middle of the orbit.
\citep[see Fig. 1 in][]{Niederste2012}.

Our goal is to explore orbital configurations or scenarios that could connect observed stellar streams A and B with the observed satellites in Malin~1, identifying progenitor candidates. Therefore, we are required to consider all detected satellite galaxies near Malin~1 (see Section \ref{sec:intro:malin1in}), and to explore the configurations or scenarios that could connect these structures in different ways. Each scenario is also evaluated using two distinct gravitational potentials for Malin~1, the NFW and the ISO DM models (Section \ref{sec:Malin1GravitationalPotential}).

For the orbital model of the satellite and stellar streams, we updated the orbital integrator code \textsc{delorean} \citep{Blana2020} with the potentials of Malin~1 and coupled it with \textsc{emcee} (Bla\~na et al., in prep.) to reconstruct the orbital trajectories of satellites while fitting the position information of the streams.
For this, we maximised the likelihood ($p = -\log(\chi^2)$) given by the chi-squared orbit fitted in projection to the stream sky coordinates, shown in Fig.~\ref{Malin1_AandBstreamPoints} and Table \ref{Table:streamPoints}. Given that the satellite-to-host mass ratio is small ($\lesssim0.1$), we model the orbits of the satellites and the streams as test particles, where $M_{\rm Satellite} / M_{\rm Malin~1} < 0.01$.
The input fixed parameters that we introduce in \textsc{delorean} are the position of Malin~1 and the positions and line-of-sight velocities ($v_{\rm los}$) of the satellites and the positions of the streams (see Table \ref{Table:streamPoints}). The free parameters in each orbit calculation are the los coordinate $Z$, the tangential velocity $V_{\rm tg}$, the position angle ${\rm PA}$, and the log posterior $ log (p)$. From the posterior sampling, we \cRB{derived} the orbital characteristics: the pericentre $R_{\rm peri}$, the apocentre $R_{\rm apo}$, and the orbital eccentricity $\epsilon$, which \cRB{helped} us classify the orbital nature and stability of each solution. More details are provided in Table \ref{tab:parameters}.

The selection of parameter ranges in each orbital scenario is based on a combination of observational constraints and iterative modelling; our general strategy involves the following steps:

\begin{itemize}
    \item[-] {Observational data:} We started the process using satellite/stream {observational data}, specifically the measured {projected sky position} and {line-of-sight velocity} for each object.
    
    \item[-] {Initial parameter space:} The orbital solution space was explored by fitting three main free parameters: the line-of-sight distance $z$, {the magnitude of the tangential velocity $V_{\rm tg}$, and the direction of $V_{\rm tg}$ is aligned with the position angle $\mathrm{PA}$, defined as the angle measured counter-clockwise from North, centred at satellite's observed position}. The ranges of initial parameters were established based on physical motivations to restrict search volume and avoid nonphysical results.
    
    \item[-] {Iterative \textsc{mcmc} refinement:} An initial \textsc{emcee} run uses wide uniform priors to sample the global parameter space. We then {iteratively refined} the prior ranges based on the most probable values (identified using the $\mathrm{\textsc{corner}}$ plots) and {repeated the \textsc{mcmc} fitting} until the posterior distributions and the resulting orbital paths {converged} to a robust solution.
    
    \item[-] {Orbit evaluation:} The resulting {3D orbital trajectories} were evaluated by checking their consistency against observed features, specifically the stream {curvature}, {direction}, and {alignment} with the disc of Malin~1.
    
    \item[-] {Best-fit solution:} The final, converged parameter distributions derived from the iterative fitting represent the {optimised orbital solution} for the stream or satellite.
\end{itemize}

\section{Results}
\label{sec:res}

\begin{table}[ht]
\centering
\caption{Malin~1 model best-fitted parameters.}
\label{tab:bestfit_params}
\tiny
\begin{tabular*}{\columnwidth}{@{\extracolsep{\fill}}llcc@{}}
\hline\hline
\multicolumn{4}{l}{I. Photometric and gas model parameters\tablefootmark{a}} \\
\hline
Component & Parameter [Unit] & \multicolumn{2}{c}{Value} \\
\hline
Bulge & $r_{\mathrm{Bulge}}$ [kpc] & \multicolumn{2}{c}{$1.0 \pm 0.1$} \\
      & $L_{\mathrm{Bulge}}$ [$10^{10} L_{\odot}$] & \multicolumn{2}{c}{$1.3 \pm 0.1$} \\
\noalign{\smallskip}
HSB disc & $z_{\mathrm{HSB}}$ [kpc] & \multicolumn{2}{c}{$0.3$\tablefootmark{i}} \\
         & $R_{\mathrm{HSB}}$ [kpc] & \multicolumn{2}{c}{$4.6 \pm 0.3$} \\
         & $L_{\mathrm{HSB}}$ [$10^{10} L_{\odot}$] & \multicolumn{2}{c}{$2.3 \pm 0.1$} \\
\noalign{\smallskip}
LSB disc & $z_{\mathrm{LSB}}$ [kpc] & \multicolumn{2}{c}{$0.3$\tablefootmark{i}} \\
         & $R_{\mathrm{LSB}}$ [kpc] & \multicolumn{2}{c}{$94^{+7}_{-6}$}\\
         & $L_{\mathrm{LSB}}$ [$10^{10} L_{\odot}$] & \multicolumn{2}{c}{$7.1^{+0.6}_{-0.5}$} \\
\noalign{\smallskip}
Gas disc & $z_{\mathrm{Gas}}$ [kpc] & \multicolumn{2}{c}{$0.3$\tablefootmark{i}} \\
         & $R_{\mathrm{Gas}}$ [kpc] & \multicolumn{2}{c}{$94^{+7}_{-6}$} \\
         & $M_{\mathrm{Gas}}$ [$10^{10} M_{\odot}$] & \multicolumn{2}{c}{$8.5 \pm 1.0$\tablefootmark{j}} \\
\hline
\multicolumn{2}{l}{II. Stellar $\gamma$:} & ISO Halo & NFW Halo \\
\hline
\multicolumn{2}{l}{$\gamma_{\mathrm{Bulge}}$ [$M_{\odot} L_{\odot}^{-1}$]} & $0.90^{+0.16}_{-0.16}$\tablefootmark{k} & $0.90^{+0.16}_{-0.16}$\tablefootmark{k} \\
\hline
\multicolumn{2}{l}{III. DM halo parameters:} & ISO Halo & NFW Halo \\
\hline
\multicolumn{2}{l}{$R_{\mathrm{vir}}$ [kpc]\tablefootmark{b}} & $282^{+14}_{-14}$ & $231^{+17}_{-15}$ \\
\multicolumn{2}{l}{$M_{\mathrm{vir}}$ [$10^{12} M_{\odot}$]\tablefootmark{c}} & $2.57^{+0.44}_{-0.38}$ & $1.41^{+0.28}_{-0.22}$ \\
\multicolumn{2}{l}{$\rho_0$ [$10^{6} M_{\odot} \mathrm{kpc}^{-3}$]\tablefootmark{d}} & $48.7 \pm 28.9$ & $6.72 \pm 4.77$ \\
\multicolumn{2}{l}{$r_c$ [kpc]\tablefootmark{e}} & $3.67^{+1.73}_{-0.85}$ & --- \\
\multicolumn{2}{l}{$r_s$ [kpc]\tablefootmark{f}} & --- & $20.2 \pm 6.6$ \\
\multicolumn{2}{l}{$c$\tablefootmark{g}} & --- & $11.4 \pm 3.8$ \\
\hline
\multicolumn{2}{l}{$\chi^2_{\mathrm{reduced}}$} & $1.21$ & $2.15$ \\
\hline
\end{tabular*}
\tablefoot{
The results include photometric and gas disk components, stellar mass-to-light ratios ($\gamma$), and Dark Matter (DM) halo parameters for both the ISO and NFW models. \\
\tablefoottext{a}{Stellar components modelled with the Miyamoto-Nagai potential (Eq. 1), where $z_h$ is the scale height, $R_d$ is the scale length, and $M_d$ is the mass.}
\tablefoottext{b}{Virial radius, $R_{200}$.}
\tablefoottext{c}{Virial mass, $M_{200}$.}
\tablefoottext{d}{Central density for the ISO model ($\rho_0$).}
\tablefoottext{e}{Bulge radius ($r_c$) for the ISO model.}
\tablefoottext{f}{Scale radius ($r_s$) for the NFW model.}
\tablefoottext{g}{Concentration parameter $c = R_{\mathrm{vir}}/r_s$ for the NFW model.}
\tablefoottext{i}{Parameter fixed during fitting.}
\tablefoottext{j}{Estimated error based on \ion{H}{i} flux uncertainty.}
\tablefoottext{k}{Fixed value based on stellar population synthesis models \citet{Lelli2010}; the quoted uncertainty is the estimated systematic error.}
}
\end{table}

Here, we first present the results of the mass and potential modelling of Malin~1. This is followed by the results of the orbital modelling of the stellar streams and satellites, considering different satellite-stream scenarios. 

\subsection{Mass and potential models for Malin~1}
\label{sec:res:potmod}

\begin{figure}[ht!]
    \centering
    \includegraphics[width=0.98\linewidth]{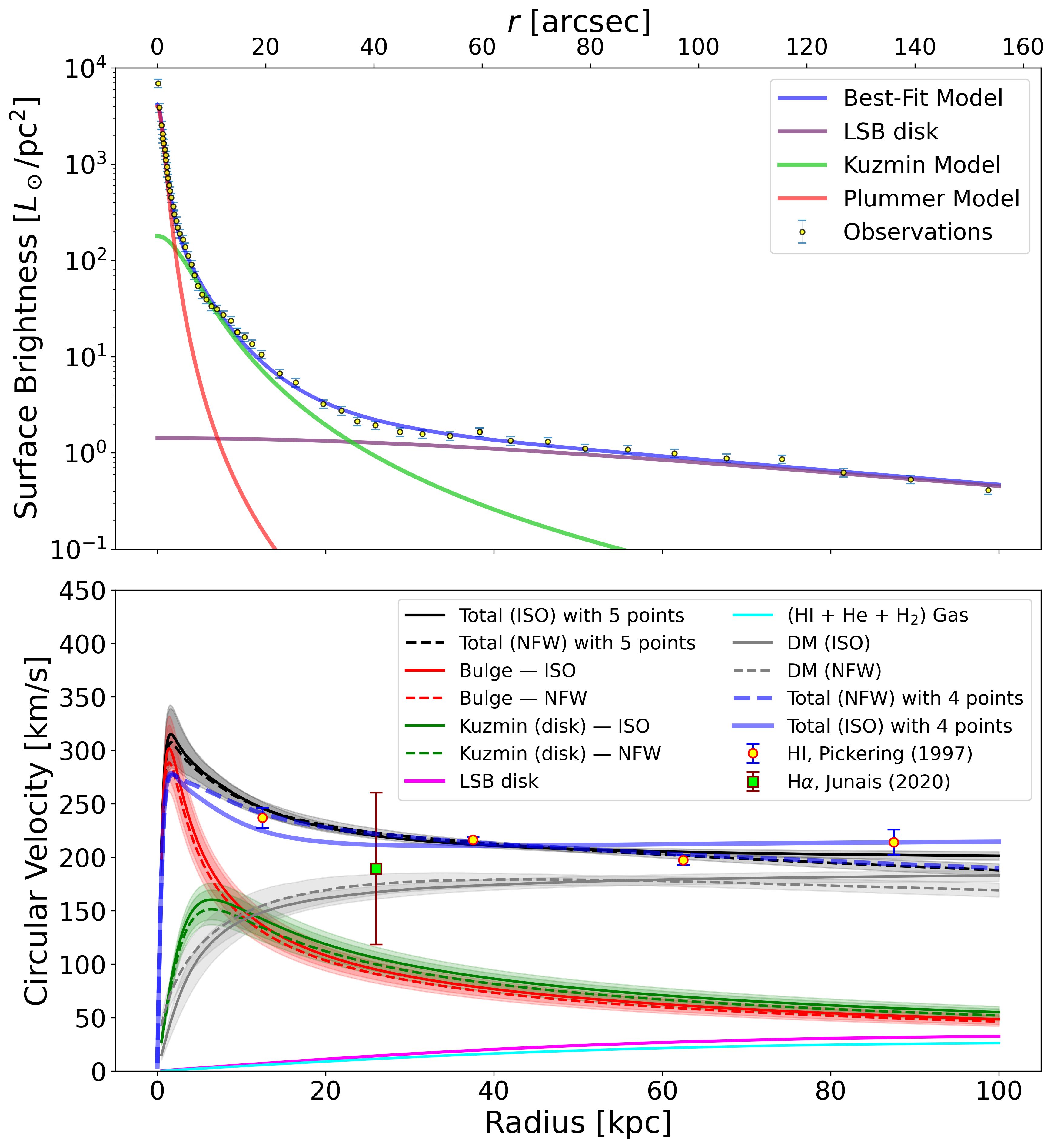}
    \vspace{-0.3cm}
    \caption{Top panel: Observed $R$-band surface brightness profile of Malin~1 as a function of the galactocentric radius \citep{Lelli2010}. Overplotted are the best-fit stellar components: a Plummer model for the bulge (red line), a Miyamoto--Nagai (Kuzmin) disc for the inner HSB component (green line), and another Miyamoto--Nagai disc for the extended LSB component (purple line). The total best-fit photometric model is shown in blue. Bottom panel: Observed H\,\textsc{i} rotation data from \citet{Pickering1997} recalculated in \citet{Lelli2010} (yellow circles) and H$\alpha$ data from \citet{Junais2020} (green square). We show the Malin~1 best model fit: the stellar bulge (red), HSB disc (green), and LSB disc (purple), along with the gaseous disc component composed of H\,\textsc{i}, He, and H$_2$ (cyan). With gray curves we show the DM halo model cases ISO (solid) and NFW (dashed). The black curves show the total mass model cases for the ISO (solid) and NFW (dashed) halos fitting five data points (four H\,\textsc{i} and one H$\alpha$). We also show a test model (blue) fitted to the H\,\textsc{i} data only that adopts a single $\gamma$ for the HSB disc and bulge. The shaded regions represent the 1$\sigma$ confidence intervals from the \textsc{mcmc} posteriors.}
    \label{fig:emceeFittings}
\end{figure}

\begin{figure}[ht!]
    \centering
    \includegraphics[width=0.98\linewidth]{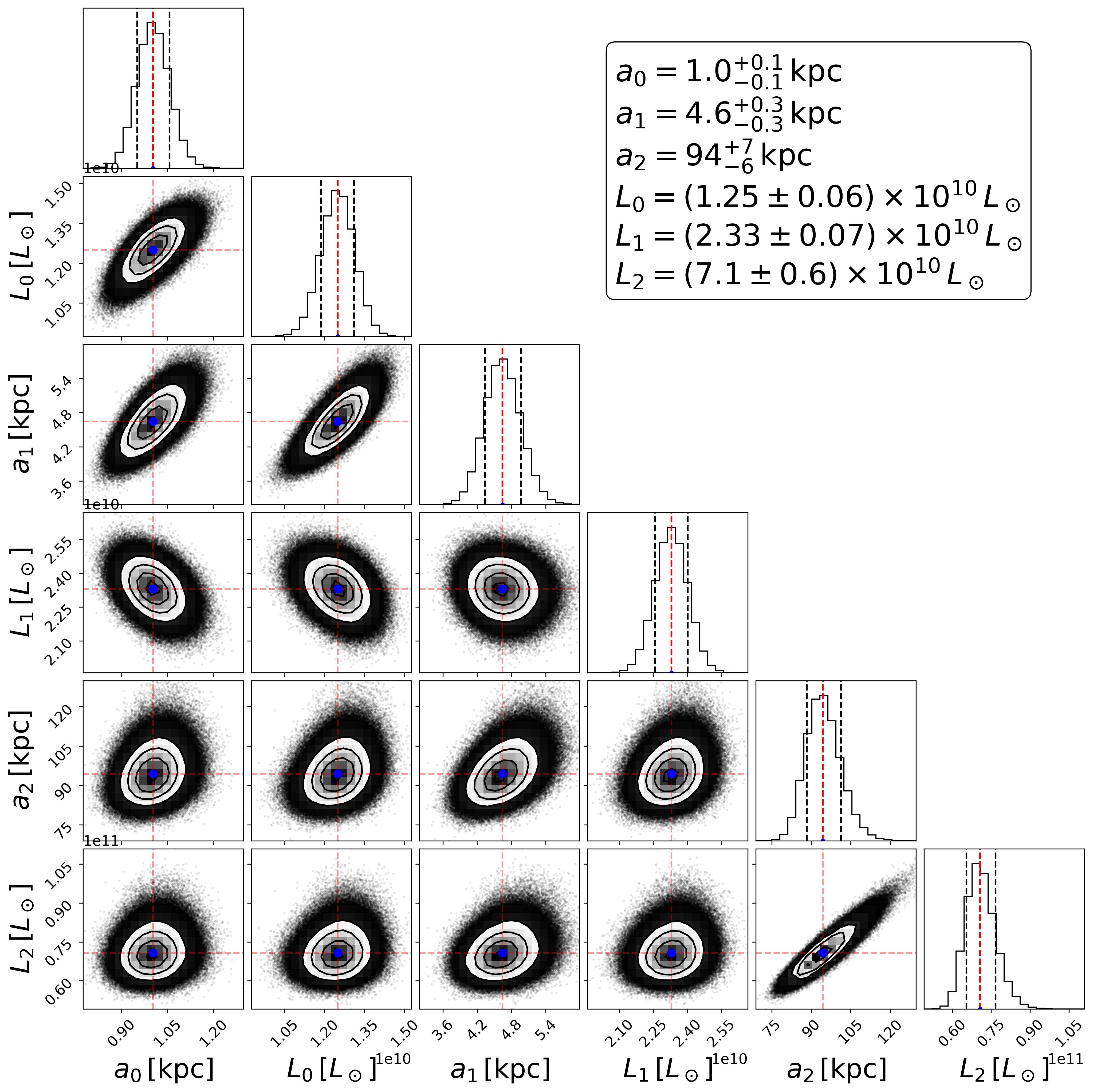} 
    \vspace{-0.2cm}
    \caption{Parameter posterior distributions for the surface brightness fitting of Malin~1.
    The scale length parameters $a_0$, $a_1$ and $a_2$, and \cRB{the} luminosities $L_0$, $L_1$, and $L_2$ are the parameters for the Bulge (Plummer), HSB disc (Kuzmin), and the LSB disc (Kuzmin) components. 
    The histograms are the \cRB{marginalised} distributions for each parameter, with the 16th, 50th, and 84th percentiles (vertical dashed lines) and values shown in the legend.}
    \label{fig:emceeSBhistograms}
    \vspace{0.2cm}
\end{figure}

{In Fig. \ref{fig:emceeFittings} (top panel), we display the surface brightness profile of the best Malin~1 luminosity model as a function of galactocentric distance. 
We also show the data points of the r-band, which were converted from the $g$ and $r$ bands \citep{Lelli2010}, revealing a good fit to the main substructures.
As expected, the fit produces a Plummer profile that characterises Malin~1's bulge well, while one Miyamoto-Nagai model (Kuzmin) represents the HSB disc, and the extended disc represents the LSB component, dominating the SB beyond 20\kpc. 
The \textsc{mcmc-emcee} posterior distributions of the luminosities and scale-length parameters are shown in Figure \ref{fig:emceeSBhistograms}. 
The components of the best fitted models are reported in Table \ref{tab:bestfit_params}, resulting in a luminosity of the inner components of $L_{\rm LSB}=3.4\times10^{10}\slu$, while the LSB component has a luminosity of $L_{\rm LSB}(R<{100\kpc})=2.2\times10^{10}\slu$ within the observed radius of $R\sim100\kpc$.}

Following this, we proceed to model the rotation curve data. For this, we used four \HI{} rotation curve data points taken from \citet{Lelli2010}. Additionally, there are recent gas kinematic observations from the central region \citep{Junais2020,Junais2024MUSEAttenuation,Johnston2024}. 
We tested the inclusion of these additional observations, considering 17 points: 10 H$\alpha$, 3 [OII], and 4 H\,\textsc{i} (see Fig.~\ref{Malin1_RCwHIHOIIHalpha}). 
However, the gas kinematics of H$\alpha$ and [OII] appear to be affected by non-circular motions, {\citep{Saburova2022aems.conf..395S, Kataria2025}}.
Including them in the fit would bias the mass models, particularly in the innermost kiloparsecs. A more sophisticated triaxial modelling of the stellar kinematics would be required; however, this would mostly improve the modelling in the central region, which is beyond the scope of this work. For this reason, and to ensure consistency with the large-scale dynamical modelling, we adopted a conservative strategy: we excluded the inner H$\alpha$ and [OII] data and retained only the outermost H$\alpha$ point, together with the four H\,\textsc{i} points. Thus, our final \cRB{rotation curve} fit is based on five data points from robust tracers (four H\,\textsc{i} + one H$\alpha$), which are less affected by non-circular motions. This approach balances the use of available kinematic constraints with the need to avoid systematics introduced by disturbed central motions. 

{The \cRB{rotation curve} modelling requires fitting a \cRB{DM} ISO or NFW halo model, and converting the luminosities to stellar mass profiles. For the latter, we implemented three independent mass-to-light ratio parameters ($\gamma$): one for the bulge, another for the HSB disc, and a third value for the LSB disc component. Although physically motivated, introducing this amount of free parameters increases the degeneracy when fitting the rotation curve.
Therefore, for our fiducial modelling we followed \citet{Lelli2010} and adopted $\gamma_{\rm LSB} = 0.90\sm\slu^{-1}$ as a fixed LSB value, and fitted $\gamma_{\mathrm{\rm Bulge}}$ and $\gamma_{\mathrm{\rm HSB}}$ for ISO halo models and NFW models. 
In addition, we chose a prior distribution strongly constrained by colour information and stellar population models from \citet{Bell2001} and other $\gamma$ constraints in Malin~1 \citep{Junais2020}.
The resulting posterior distribution of the models that used the ISO halo is shown in Fig.~\ref{fig:ThreeMassToLightRatiosForISO}, and NFW in Fig. \ref{fig:ThreeMassToLightRatiosForNFW}.
We also tested the degeneracy of $\gamma_{\rm LSB}$ by fixing the value of $\gamma_{\rm Bulge} \approx 3.8\sm\slu^{-1}$, $\gamma_{\rm HSB} \approx 1.5\sm\slu^{-1}$ based on previous work \citep[][see their Fig.~9]{Junais2020}, leaving $\gamma_{\rm LSB}$ as a free parameter, obtaining a range $\simeq 0.86$--$0.90\sm\slu^{-1}$ in concordance with \citet{Lelli2010}.
Furthermore, we also tested using a mass-to-light value for the bulge and disc region $\gamma_{\mathrm{\rm Bulge}}=\gamma_{\mathrm{\rm HSB}}=3.4\sm\slu^{-1}$, finding results consistent with \citet{Lelli2010}.
The best parameters are reported in Table~\ref{tab:bestfit_params}, showing $\gamma$ for each stellar component, along with the corresponding DM halo parameters for both the ISO and NFW models.}

{In general, we find that the bulge region has systematically higher $\gamma$ values than the HSB disc component.
The rotation curves of the best ISO and NFW models are shown in Figure \ref{fig:emceeFittings} (bottom panel).
The decomposition also shows the contribution of different mass components: stellar bulge, HSB/LSB discs, and the gaseous disc (\HI{}, He and $H_2$). 
Both the high- and low-surface brightness (HSB and LSB) regions of the disc are considered. The figure further compares the performance of the two halo \cRB{parametrizations}, highlighting the differences between the NFW and ISO models.
Figure \ref{fig:emceeSBhistograms} illustrates the optimi\cRB{s}ed \textsc{emcee} fit parameters for the surface brightness model, while Figure \ref{fig:ThreeMassToLightRatiosForISO} shows the equivalent parameters for the rotation curve model. The data compiled for the two profiles is presented in Table \ref{tab:bestfit_params}. 
The model with the ISO \cRB{DM} profile provides better fits of the rotation curve, as shown by the reduced ($\chi^2$) of the \textsc{mcmc}. 
This is due to the flat rotation curve behaviour of the isothermal profile, while the NFW halo drops faster at the same distances and similar masses. Moreover, we also find that the models with NFW produce higher values of $\gamma$ than the ISO models. This is due to the cuspy nature of the NFW dark halo model, which introduces more mass in the centre, forcing the stellar mass to have lower $\gamma$. This results in NFW halos having virial masses of $1.4\times10^{12}\sm$ and lower than the ISO models with $2.6\times10^{12}\sm$. Furthermore, our test with equal central $\gamma$ result in more massive ISO halo models with $3.7\times10^{12}\sm$.
}

{Given that the differences between the potential models with ISO and NFW haloes are  larger than the differences within each halo model profile, we proceed to use both sets of models to consider the uncertainties in the potential that are used to explore the orbital scenarios in the following section.}

\begin{figure}[ht!]
    \centering
    \includegraphics[width=0.95\linewidth]{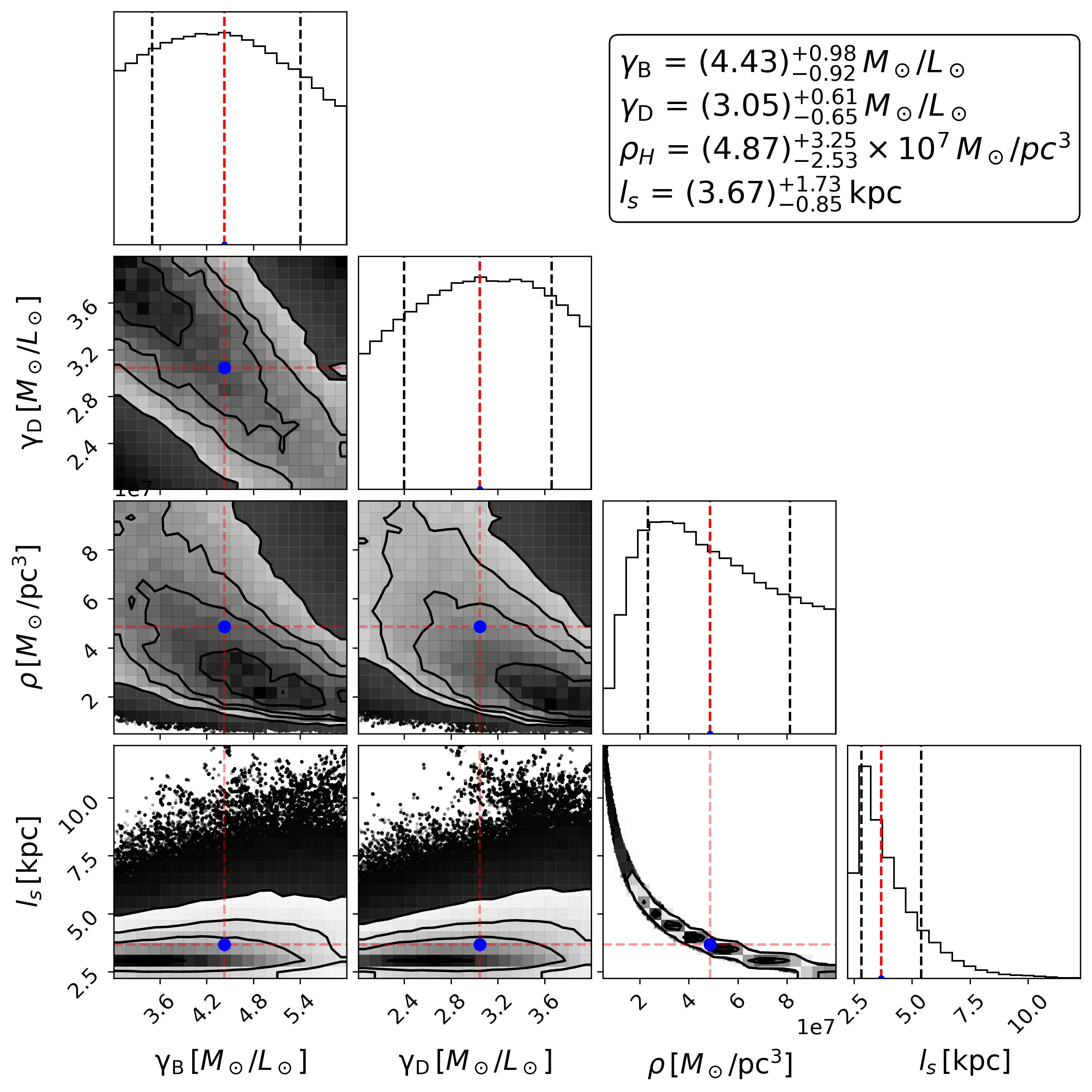}
    \vspace{-0.2cm}
    
    \caption{Posterior distributions of the mass-to-light ratios for the bulge, HSB disc, and the ISO DM halo parameters: central density ($\rho_{\rm Halo}$) and the core radius ($r_c$). Diagonal panels show the marginali\cRB{s}ed distributions with the 16th, 50th, and 84th percentiles indicated by dashed lines. Contour plots display parameter covariances. The legend reports the best-fit values with $1\sigma$ uncertainties. The ISO model provides robust fits, with maximum-a-posteriori values (blue points) close to the medians.}
    \label{fig:ThreeMassToLightRatiosForISO}
\end{figure}

\begin{figure*}[ht!]
    \centering
    \includegraphics[width=0.45\linewidth]{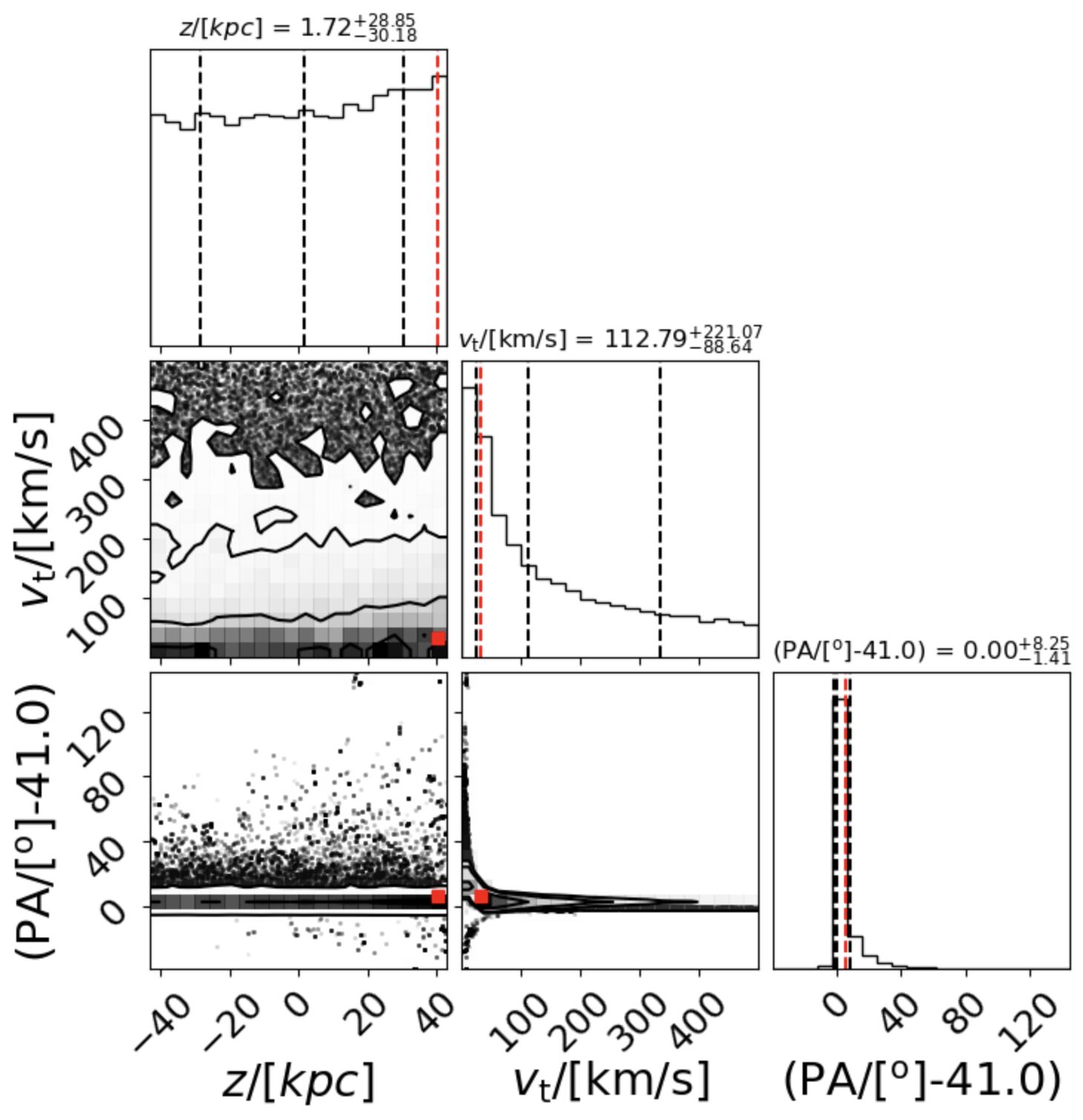}
    \hspace{0.05\linewidth}
    \includegraphics[width=0.45\linewidth]{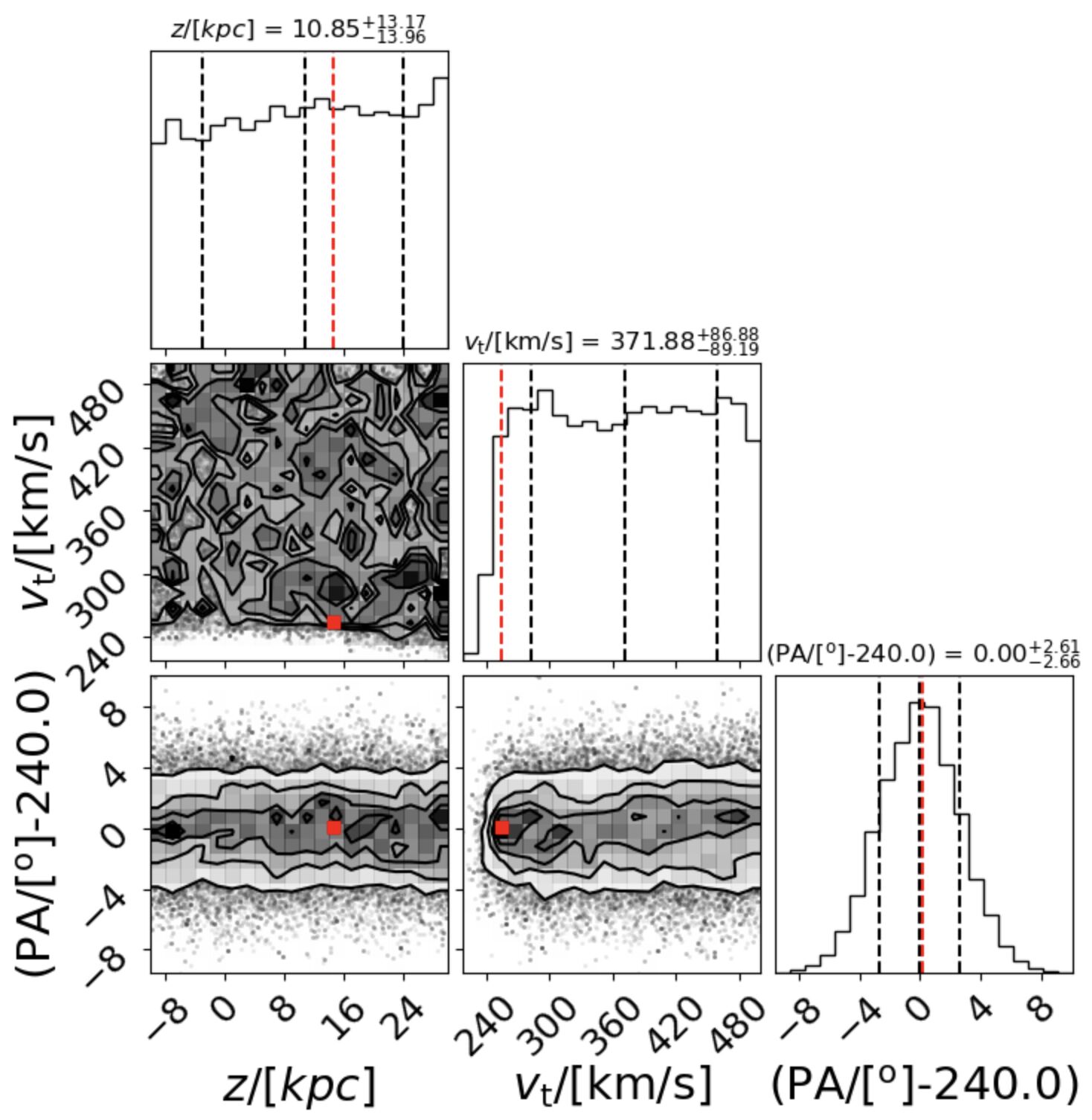}
    
\caption{Corner plots of the posterior distributions of orbital parameters ($z$, $v_t$, and $\mathrm{PA}$ offset) obtained with \texttt{emcee}. {Left:} Scenario~I-a (radial orbit, ISO halo). {Right:} Scenario~IV-a (polar orbit, ISO halo). Diagonal panels show marginali\cRB{s}ed 1D distributions with dashed lines marking the median and $1\sigma$ uncertainties. Off-diagonal panels show 2D covariances with $68\%$ and $95\%$ confidence regions. These two scenarios illustrate contrasting orbital families (radial \cRB{versus} \ polar) within the {ISO} halo parametrization.}
\label{fig:CornerPlots_ISO}

\end{figure*}

\subsection{Scenarios connecting possible progenitor satellite candidates and stellar streams}
\label{sec:res:scen}

Our goal is to search for orbital solutions of satellite galaxies that can overlap with the stellar streams.
From these we focus on radial orbits that reach the central regions of Malin~1 where the satellites could have experienced the most intense tidal forces. 
We also considered polar orbital solutions, similar to those observed in the Sombrero galaxy \citep{Martinez-Delgado2021}, where the satellite has near circular orbits while being tidally stripped. 

{We defined five scenarios that depend on what satellite is assumed to be the progenitor, and two subcases (a and b) that depend on what combinations of the streams are assumed to be associated with each satellite. 
These are summari\cRB{s}e in Table \ref{table:ScenariosUnified}.
For many scenarios and cases, we found radial and polar orbital solutions.
In addition, we tested two different DM potentials (NFW and ISO) for each scenario.}

{Following our procedure to constrain the regions where the best orbits are found (see Section \ref{sec:orbitalmodelling}), we converge to the parameter ranges reported in 
Table~\ref{tab:scenarioranges} and Table~\ref{tab:scenarioranges2} where we justify each. Table \ref{tab:DMcomparison} reports the Best-fit posterior probabilities ($p$) for all the scenario models.
The investigation consists of millions of orbits per scenario, resulting in a comprehensive analysis of $\sim130$ million orbits in total to explore the parameter space.}
Here, we highlight the solutions that most accurately correspond to the observational data and adhere to physical constraints, such as having bounded orbital solutions. However, unbound solutions were also found.
Here, we proceed to describe each scenario in detail:

\begin{table}[ht]
\centering
\caption{Scenarios defining orbital configurations.}
\label{table:ScenariosUnified}
\tiny
\begin{tabular*}{\columnwidth}{@{\extracolsep{\fill}}ccc@{}}
\hline\hline
Scenario & Case & Combinations \\
\hline
I  & a & {eM1} + {sB} \\
   & b & {eM1} + {sA} + {sB} \\
\noalign{\smallskip}
II & a & {M1C} + {sB} \\
   & b & {M1C} + {sA} + {sB} \\
\noalign{\smallskip}
III & a & {M1B} + {sB} \\
    & b & {M1B} + {sA} + {sB} \\
\noalign{\smallskip}
IV & a & {M1A} + {sA} \\
   & b & {M1A} + {sA} + {sB} \\
\noalign{\smallskip}
V  & a & {sA} \\
   & b & {sB} \\
\hline
\end{tabular*}
\tablefoot{
Scenarios defining orbital configurations that connect a specific satellite galaxy as the progenitor candidate for a single or both streams (cases a and b). Scenario V assumes that the stream is the disrupted progenitor. Each scenario has been evaluated using two halo models (NFW and ISO). \\
{Abbreviations:} M1 = Malin~1, sA = Stream A, sB = Stream B, eM1 = Exo-Malin~1, M1A = Malin~1A, M1B = Malin~1B, M1C = Malin~1C.
}
\end{table}

In the following, we discuss each scenario and the main orbital properties obtained from the modelling. 
{In general, we managed to identify orbital solutions that have a radial behavio\cRB{u}r, and orbits with a more circular morphology that due to the satellite configurations produce polar-like orbital solutions perpendicular to Malin~1 LSB dis\cRB{c}. 
Furthermore, we identified scenarios where the orbital solutions that are gravitationally bound (I-a, II-a, II-b, IV-a, and V-a), and unbound (I-b, III-a, III-b, IV-b, and V-b).
These correspond to the following:}

\begin{itemize}
\item Scenario I-a (Radial and Polar): {eM1} + {sB}. \\ We constrain the current tangential velocity of the candidate progenitor {eM1} by computing various orbits backward in time. The  stellar stream {sB} appears spatially connected to the central region of {M1} and the offset position of {eM1} (as initially noted by its position relative to labels 3 and 4 in Fig.~\ref{Fig:Malin1EnvironmentIn}). Figure \ref{fig:CornerPlots_ISO} left panel illustrates the posterior distributions for the orbital parameters of {Scenario I-a}, showing the $1\sigma$ uncertainties derived from the \textsc{mcmc} samples.
Figure \ref{fig:SceIarSceIp} illustrates two representative solutions of Scenario I-a: (a) a radial orbit (using the ISO DM potential) and (b) a polar orbit (using the NFW DM potential). The uncertainty bands ${1\sigma}$  shown in the temporal and kinematic evolution panels (iii and vi) were derived from 300 \textsc{mcmc} samples. {Radial Orbit:} The best-fit radial orbit, integrated over $[-13.0, 20.0]\ \text{Gyr}$, demonstrates a close passage consistent with the formation of sB as the trailing tidal segment of eM1. The pericentre passage occurred at ${P \approx 33\ \text{kpc}}$ at ${T_{\text{p}} \approx -1.6\ \text{Gyr}}$, with a velocity of ${|V_{\text{p}}| \approx 548\ \text{km/s}}$. This orbit yields a large line-of-sight velocity variation ($\Delta V_{\text{LoS}} \approx 650\ \text{km/s}$), which is consistent with the velocity spread expected for a plunging orbit. 
{This solution reveals that eM1 could be a backsplash satellite with a long orbital period.
N-body simulations have shown that depending on satellite and stream location along the orbit, the leading or trailing streams can be stretched or diluted along the orbit producing apparent gaps between the satellite and the stream \citep[e.g.][]{Smith2013}. 
Future follow-up deeper observations and simulations will further investigate this scenario.}
{Polar Orbit:} The best-fit polar orbit, utili\cRB{s}ing the NFW potential, links eM1 and sB via a more distant pericentre: ${P \approx 147\ \text{kpc}}$ at ${T_{\text{p}} \approx -2.4\ \text{Gyr}}$, with ${|V_{\text{p}}| \approx 318\ \text{km/s}}$. Despite the larger pericentre, the orbit's alignment is still consistent with sB being the trailing component of eM1. The lower $\Delta V_{\text{LoS}} \approx 350\ \text{km/s}$ is characteristic of the shallower tidal interaction experienced in the cuspy NFW potential. Both Scenarios I-a Radial and Polar, are statistically viable solutions, with Scenario I-a Polar offering a higher posterior probability, as discussed in detail in Section \ref{sec:comparisons}.
\end{itemize}

\begin{figure*}[t]
   \centering
   \begin{subfigure}{\textwidth}
      \centering
      \includegraphics[width=0.80\linewidth]{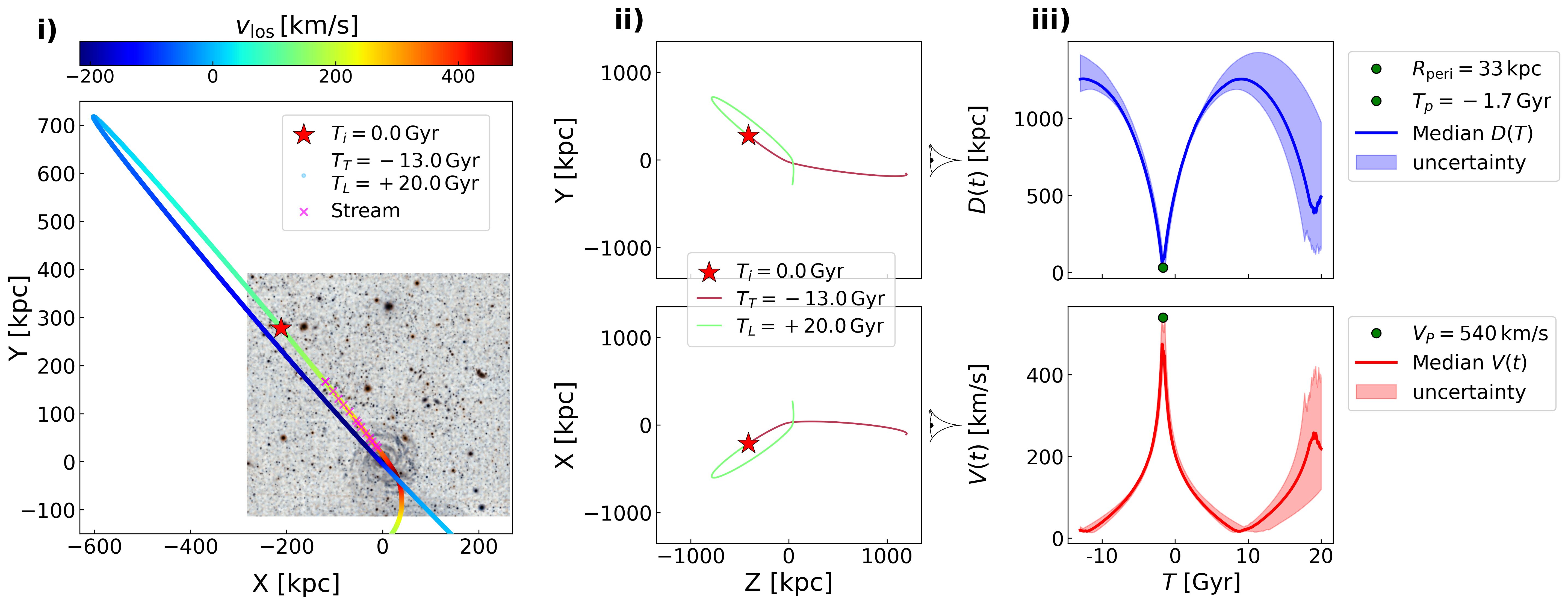}
      \caption{Scenario I-a: eM1+sB (radial)}
   \end{subfigure}
   \vspace{0.3cm}

   \begin{subfigure}{\textwidth}
      \centering
      \includegraphics[width=0.80\linewidth]{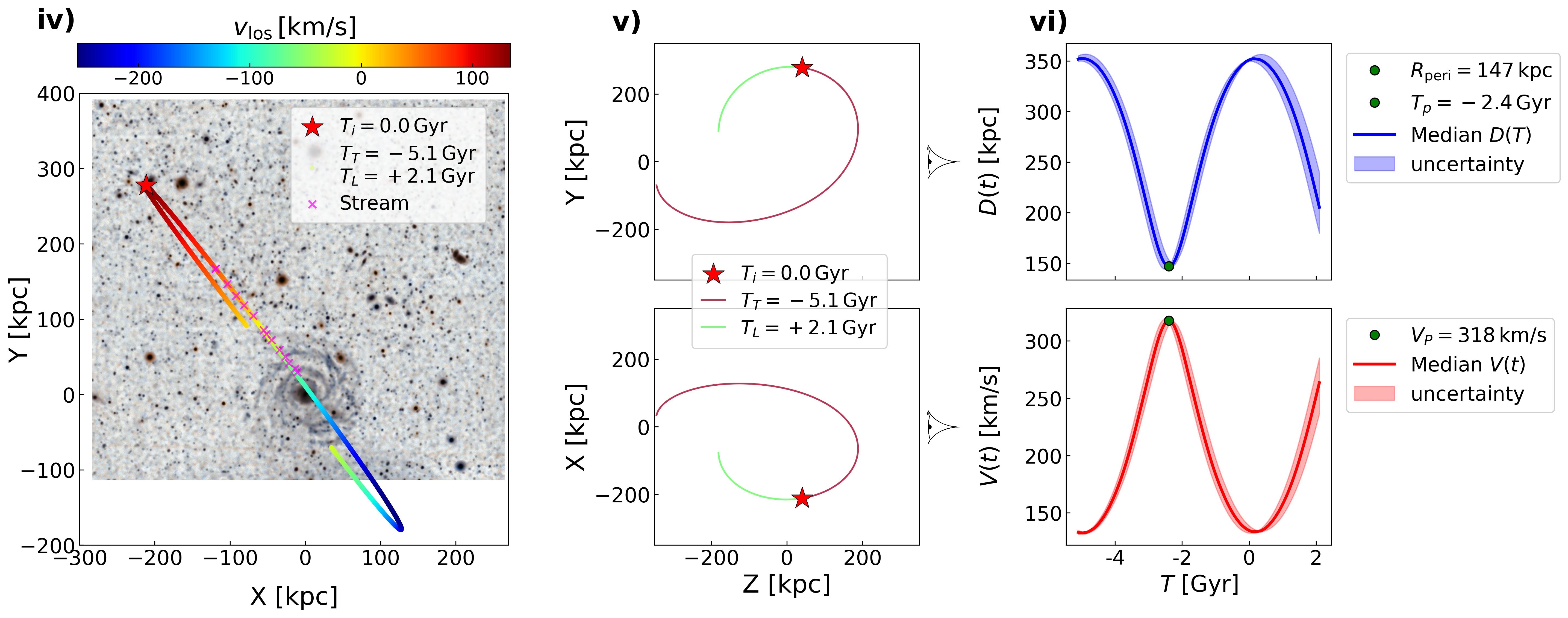}
      \caption{Scenario I-a: eM1+sB (polar)}
   \end{subfigure}
    \caption{
    Best-fit integrated orbits for the eM1+sB model, comparing the two highest-likelihood orbit classes: (a) Scenario I-a (radial) and (b) Scenario I-b (polar). The uncertainty bands in panels (iii) and (vi) were derived from the $300$ \textsc{mcmc} samples with the highest posterior values. 
    {Spatial projections (Panels i and iv):} The integrated orbit (colo\cRB{u}red line) is overlaid on the optical image of Malin~1 and its environment \citep{Galaz2015}. The orbit is colour-coded by the line-of-sight velocity ($v_{\text{los}}$). The red star marks the progenitor (eM1) at $T_i = 0.0\ \text{Gyr}$ (present), and the magenta cross denotes stellar stream markers (sB). The $T_{\text{T}}$ (Trailing) and $T_{\text{L}}$ (Leading) stream times are marked with dots along the orbit. {Orbital projections (Panels ii and v):} The Z-Y and Z-X orbital planes are shown for the orbits. The origin $X,Y = (0, 0)\ \text{kpc}$ is the Malin~1 centre. The observer is conceptually located to the right (positive Z and X directions, as indicated by the eye icon). The legends indicate the Trailing ($T_{\text{T}}$) and Leading ($T_{\text{L}}$) integration times and colours used for the orbit segments. {Temporal and kinematic evolution (Panels iii and vi):} The top sub-panels show the orbital distance $D(t)$ (blue) and the bottom sub-panels show the velocity $V(t)$ (red) as a function of time $T$. The solid line is the median orbit, and the shaded regions indicate the $1\sigma$ uncertainty. The green circle marks the maximum a posteriori pericentre (distance $R_{\mathrm{peri}}$ and time $T_p$) and pericentre velocity ($V_p$), determined as the median of the extrema across all \textsc{mcmc} samples. {Scenario-specific parameters:} {I-a (radial, ISO DM):} Best orbit integrated over $[-13.0, 20.0]\ \text{Gyr}$. Pericentre passage: ${R_{\mathrm{peri}} \approx 33\ \text{kpc}}$ at ${T_{\text{p}} \approx -1.6\ \text{Gyr}}$, with ${|V_{\text{p}}| \approx 540\ \text{km/s}}$. Max $\Delta V_{\text{LoS}} \approx 650\ \text{km/s}$. $z \sim -412$ kpc, $v_t \sim 153$ km/s, $\mathrm{PA} \sim 40^\circ$, apocentre $R_{\text{apo}} \sim 1194\ \text{kpc}$ and eccentricity $\epsilon \sim 0.68$. This orbit is consistent with sB being the trailing part of eM1. {I-b (polar, NFW DM):} Best orbit integrated over $[-5.1, 2.1]\ \text{Gyr}$. Pericentre passage: ${R_{\mathrm{peri}} \approx 147\ \text{kpc}}$ at ${T_{\text{p}} \approx -2.4\ \text{Gyr}}$, with ${|V_{\text{p}}| \approx 318\ \text{km/s}}$. Max $\Delta V_{\text{LoS}} \approx 350\ \text{km/s}$. $z \sim -40$ kpc, $v_t \sim 33$ km/s, $\mathrm{PA} \sim 47^\circ$, apocentre $R_{\text{apo}} \sim 354\ \text{kpc}$ and eccentricity $\epsilon \sim 0.09$. This orbit, which assumes an NFW DM profile, is also consistent with sB being the trailing part of eM1.
    }
   \label{fig:SceIarSceIp}
\end{figure*}

\begin{figure*}[t]
\centering

    \begin{subfigure}{\textwidth}
        \centering
        \includegraphics[width=0.9\linewidth]{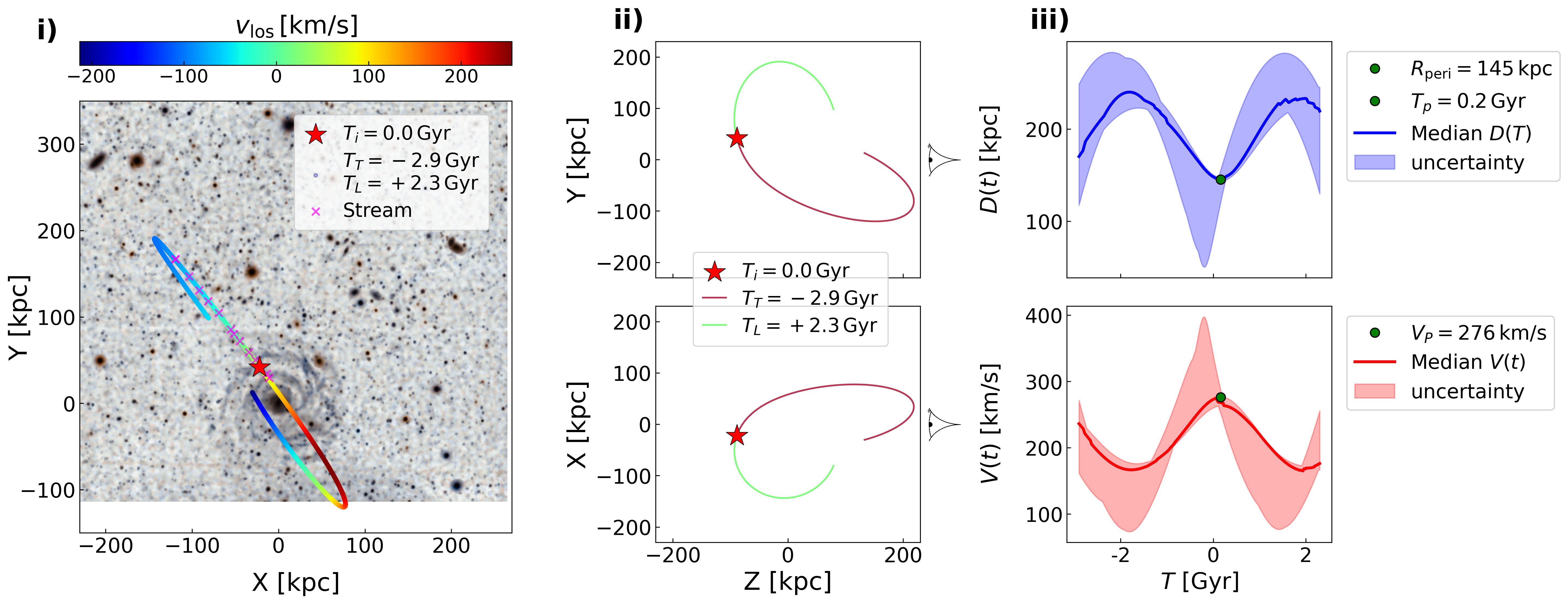}
        \caption{Scenario II-a: M1C + sB (polar)}
    \end{subfigure}
    
    \vspace{0.3cm}
    
    \begin{subfigure}{\textwidth}
        \centering
        \includegraphics[width=0.9\linewidth]{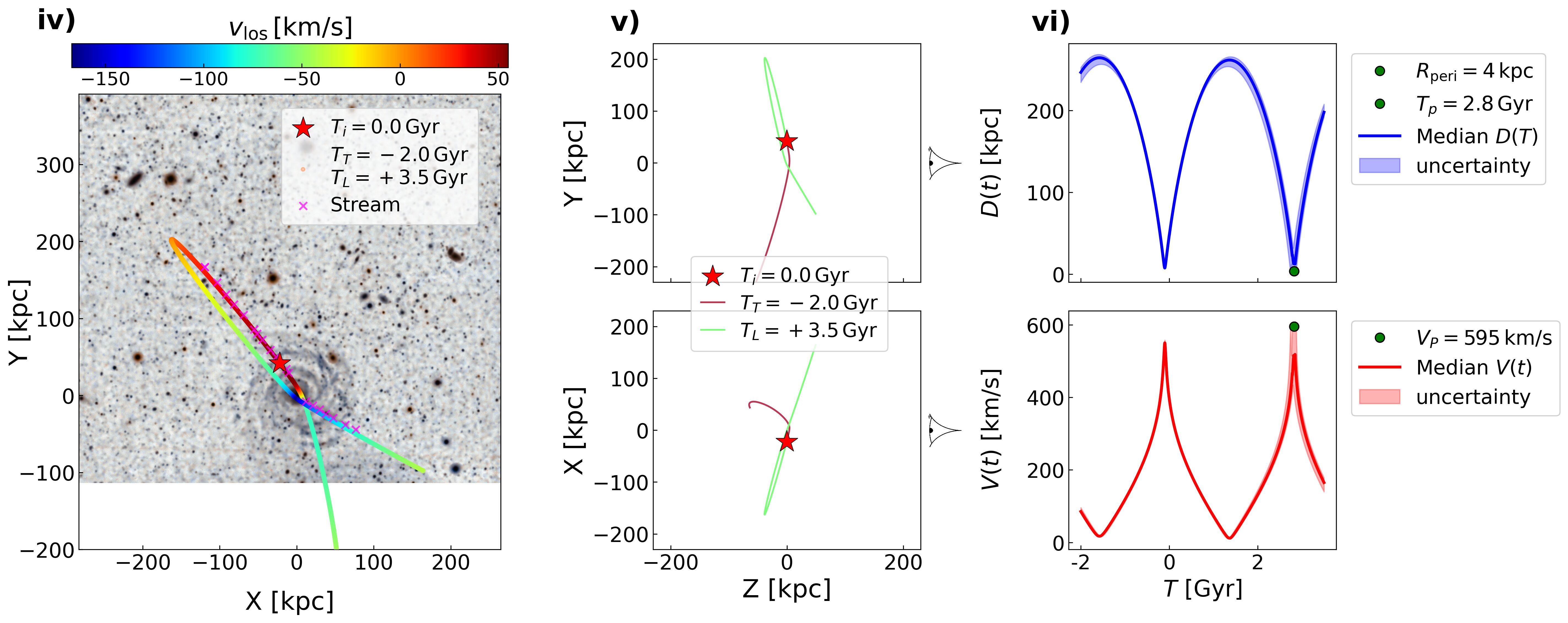}
        \caption{Scenario II-b: M1C + sA + sB (radial)}
    \end{subfigure}

\caption{
Best-fit integrated orbits for the progenitor {M1C} and two scenarios. {Panel (a) corresponds to Scenario II-a} (${M1C}+{sB}$) in panels i), ii), and iii). {Panel (b) corresponds to Scenario II-b} (${M1C}+{sA} + {sB}$) in panels iv), v), and vi). The uncertainty bands (iii, vi) use the $300$ highest-posterior \textsc{mcmc} samples.
{Panels i and iv (Spatial):} Integrated orbit (colour-coded by $v_{\text{los}}$) overlaid on Malin~1's optical image. The red star is the progenitor ({M1C}) at $T_i=0$, and the magenta cross marks stream components. {Panels ii and v (Orbital):} $Z-Y$ and $Z-X$ orbital planes. The observer is to the right (positive $Z$ and $X$). {Panels iii and vi (Evolution):} Top: orbital distance $D(t)$ (blue); Bottom: velocity $V(t)$ (red). The green circle marks the maximum a posteriori pericentre ($R_{\mathrm{peri}}$, $T_p$) and velocity ($V_p$) from the \textsc{mcmc} sample extrema.
{Scenario-specific parameters:}
{II-a (M1C+sB, radial, ISO DM):} Integrated over $[-2.9, 2.3]\ \text{Gyr}$. ${R_{\mathrm{peri}} \approx 145.0\ \text{kpc}}$ at ${T_{\text{p}} \approx 0.2\ \text{Gyr}}$, ${|V_{\text{p}}| \approx 276\ \text{km/s}}$. Apocentre $R_{\text{apo}} \sim 250\ \text{kpc}$ and eccentricity $\epsilon \sim 0.82$. This solution is ${unbound}$ and dynamically disfavo\cRB{u}red.
{II-b (M1C+sA+sB, radial, ISO DM):} Integrated over $[-2.0, 3.5]\ \text{Gyr}$. ${R_{\mathrm{peri}} \approx 3.0\ \text{kpc}}$ at ${T_{\text{p}} \approx 2.8\ \text{Gyr}}$, ${|V_{\text{p}}| \approx 598\ \text{km/s}}$. $R_{\text{apo}} \sim 1530\ \text{kpc}$ and eccentricity $\epsilon \sim 0.56$. The leading pericentre passage occurred $2.9$ Gyr ago (first passage) and $0.33$ Gyr ago (leading arm/second passage). Significantly, this is the only solution found that dynamically links two separate streams (sA and sB) via a common progenitor (M1C).
}
\label{fig:SceIIb}
\end{figure*}

\begin{figure*}[t]
   \centering
   \begin{subfigure}{\textwidth}
      \centering
      \includegraphics[width=0.80\linewidth]{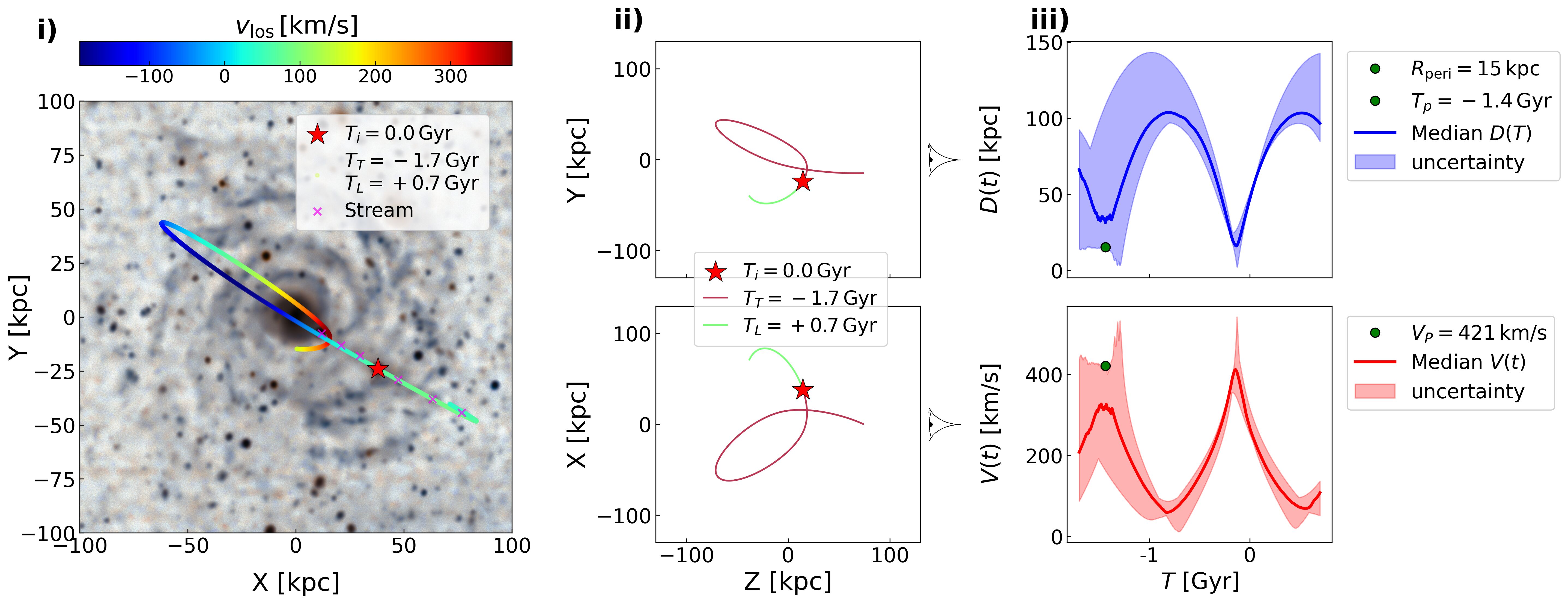}
      \caption{Scenario IV-a: M1A + sA (polar)}
   \end{subfigure}

   \vspace{0.3cm}

   \begin{subfigure}{\textwidth}
      \centering
      \includegraphics[width=0.80\linewidth]{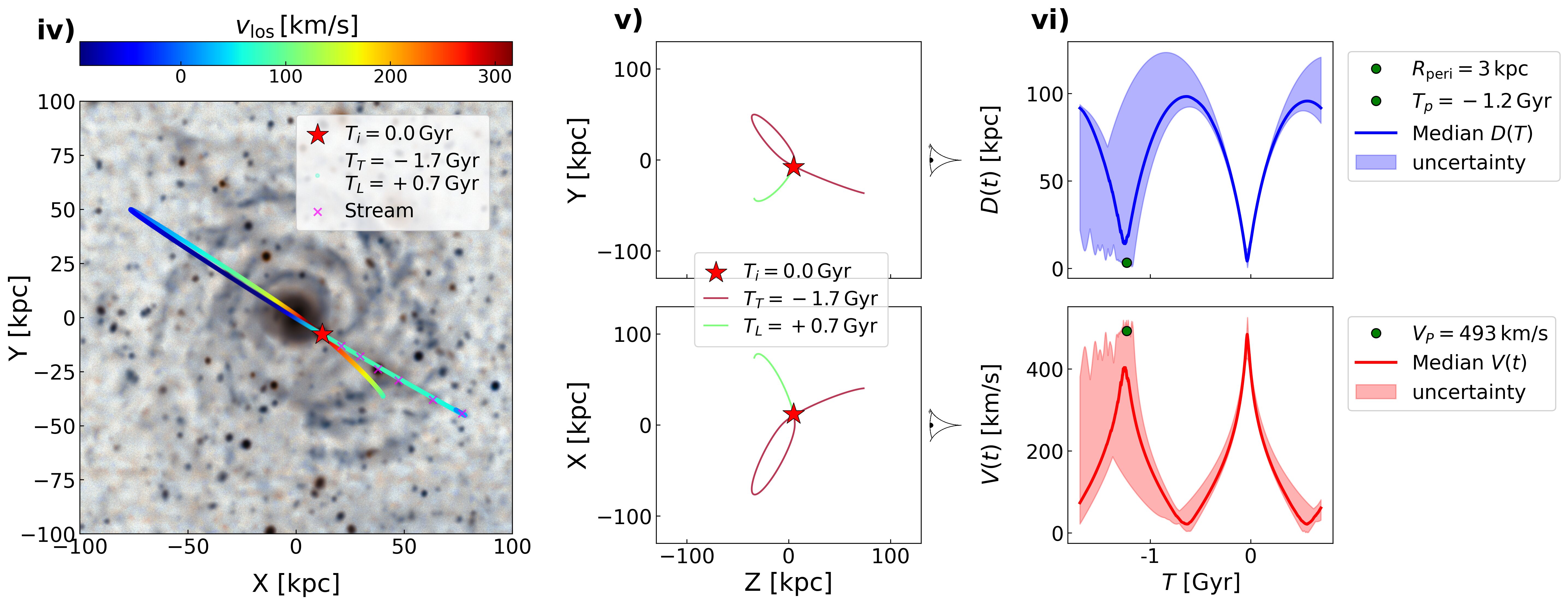}
      \caption{Scenario V-a: sA only, progenitor destroyed (radial)}
   \end{subfigure}

\caption{Best-fit integrated orbits for the stream sA, comparing the two most likely scenarios: (a) Scenario IV-a (M1A progenitor, ISO DM) and (b) Scenario V-a (sA progenitor destroyed, NFW DM). The ${1\sigma}$ uncertainty bands in panels (iii) and (vi) were derived from $300$ \textsc{mcmc} samples. {Spatial projections (Panels i and iv):} \cRB{Optical image of Malin 1 with the} integrated orbit (colo\cRB{u}red line) overlaid. The orbit is colour-coded by the line-of-sight velocity ($v_{\text{los}}$). The red star marks the progenitor (M1A/sA) at $T_i = 0.0\ \text{Gyr}$ (present), and the magenta cross denotes the stellar stream (sA). The $T_{\text{T}}$ (Trailing) and $T_{\text{L}}$ (Leading) stream times are marked with dots. {Orbital projections (Panels ii and v):} Z-Y and Z-X orbital planes. The origin $X,Y = (0, 0)\ \text{kpc}$ is the Malin~1 centre. The observer is conceptually located to the right (positive Z and X directions, as indicated by the eye icon). The legends indicate the Trailing ($T_{\text{T}}$) and Leading ($T_{\text{L}}$) integration times and \cRB{the} colours used for the orbit segments. {Temporal and kinematic evolution (Panels iii and vi):} \cRB{O}rbital distance $D(t)$ (\cRB{in} blue\cRB{; top sub-panels}) and velocity $V(t)$ (\cRB{in} red\cRB{;bottom sub-panels}) as a function of time $T$. The solid line is the median orbit, and the shaded regions indicate the $1\sigma$ uncertainty. The green circle marks the maximum a posteriori defined pericentre (distance $R_{\mathrm{peri}}$ and time $T_p$) and pericentre velocity ($V_p$), determined as the median of the extrema across all \textsc{mcmc} samples. {Scenario-specific parameters:} {IV-a (M1A+sA, polar, ISO DM):} Best orbit integrated over $[-1.7, 0.7]\ \text{Gyr}$. Pericentre passage: ${R_{\mathrm{peri}} \approx 15\ \text{kpc}}$ at ${T_{\text{p}} \approx -1.4\ \text{Gyr}}$, with ${|V_{\text{p}}| \approx 421\ \text{km/s}}$. Max ${\Delta V_{\text{LoS}} \approx 400\ \text{km/s}}$. This represents a tight polar solution for the M1A progenitor. {V-a (sA only, radial, NFW DM):} Best orbit integrated over $[-1.7, 0.7]\ \text{Gyr}$. Pericentre passage: ${R_{\mathrm{peri}} \approx 3\ \text{kpc}}$ at ${T_{\text{p}} \approx -1.2\ \text{Gyr}}$, with ${|V_{\text{p}}| \approx 493\ \text{km/s}}$. Max ${\Delta V_{\text{LoS}} \approx 400\ \text{km/s}}$. This model shows a very close radial passage that leads to the complete destruction of the progenitor (sA). The proximity of the solutions (same time frame, similar velocities \cRB{and} geometry) suggests that the orbital characteristics of sA are strongly constrained by the streams regardless of the progenitor's current status (M1A intact \cRB{versus} sA destroyed).}
   \label{fig:SceIVaSceVa}
\end{figure*}

\begin{itemize}
    \item Scenario I-b: {eM1} + {sB} + {sA}. \\ In our search for a radial orbit that could simultaneously fit both streams, a solution was not possible. Instead, we found a highly eccentric orbit that apparently is {unbound} from {Malin~1}. The pericentric passage is elevated, occurring at ${694\ \text{kpc}}$ with a velocity of ${933\ \text{km/s}}$, which happened ${500\ \text{Myr}}$ ago. This high-energy orbit (shown in panel i) of Fig. \ref{fig:notFits}) suggests that the progenitor {eM1} is currently escaping the gravitational potential of {Malin~1}.
\end{itemize}

\begin{itemize}
    \item Scenario II-a: {M1C} + {sB}. \\ This scenario explores {M1C} as the progenitor of {sB}, suggesting that {sB} contains stellar remnants and potentially undetected gas once linked to {M1C}. The probable orbit, shown in panel (a) Fig. \ref{fig:SceIIb} and integrated over $[-2.9, 2.3]\ \text{Gyr}$, experienced its pericentre passage approximately ${0.2\ \text{Gyr}}$ ago. {Key Finding:} The orbit is {bound} (as seen in the distance-time plot) and polar to {Malin~1}. It exhibits a moderate dynamic interaction, with line-of-sight velocities reaching ${\Delta V_{\text{los}} \approx 400\ \text{km/s}}$. This bound outcome is in agreement with the successful bound solution found in Scenario {II-b}.
\end{itemize}

\begin{itemize}
\item Scenario II-b: {M1C} + {sA} + {sB}. \\
This setup explores a crucial dynamical link: the possibility that both stellar streams, {sA} and {sB}, may be composed of stellar remnants originating from a single progenitor, the companion {M1C}. This model assumes a highly tidally disruptive system where the leading and trailing arms of the progenitor connects both streams, as suggested by previous numerical studies of satellite galaxy tidal evolution \citep[e.g.][]{Niederste2012}. Figure \ref{fig:SceIIb} shows the best-fit radial orbit solution, integrated over $[-2.0, 3.5]\ \text{Gyr}$ using the preferred {ISO} DM potential. {Key Findings:} This is the {only solution found} that successfully links two separate streams ({sA} and {sB}) via a common progenitor ({M1C}). The orbit is highly radial and plunges deep into the Malin~1 potential well. The model predicts an extremely close {pericentre passage at $P \approx 3\ \text{kpc}$}, occurring ${2.8\ \text{Gyr}}$ in the leading component ($T_{\text{L}}$). The current position of {M1C} places it on this highly energetic orbit, with a predicted pericentre velocity of ${V_{\text{p}} \approx 598\ \text{km/s}}$. The temporal evolution (Fig. \ref{fig:SceIIb}, panel iii) indicates that the stream {sA} is linked to an earlier passage of {M1C} approximately ${0.33\ \text{Gyr}}$ ago, while stream {sB} is associated with the subsequent passage. This scenario requires significant tidal stripping due to the very small pericentre radius.
\end{itemize}

\begin{itemize}
\item Scenario III-a: {M1B} + {sB}. \\ We investigate the possibility that {M1B} is the progenitor of the stellar stream {sB}. The best-fit orbit, shown in Panel iv), Fig. \ref{fig:notFits}, suggests a highly eccentric association ($\epsilon \sim 0.88$) with a pericentre of ${P \sim 7\ \text{kpc}}$ and a velocity of $|V| \sim 707\ \text{km/s}$ occurring $\sim 0.01\ \text{Gyr}$ ago. {Key Finding:} This close, high-velocity encounter is dynamically possible but results in an ${unbound}$ orbit, making it dynamically disfavo\cRB{u}red as the progenitor. 
\end{itemize}

\begin{itemize}
    \item Scenario III-b: {M1B} + {sB} + {sA}. \\ This situation seeks {M1B} as the progenitor \cRB{of} both {sB} and {sA} simultaneously. {Key Finding:} It was not possible to find a single fitting orbit to connect both streams, and the best solution found was an apparently {unbound} orbit, which is shown in panel ii) of Fig. \ref{fig:notFits}. The pericentre passage occurred relatively recently, ${20\ \text{Myr}}$ ago, at ${32\ \text{kpc}}$ with a very high velocity of ${\sim 705\ \text{km/s}}$.
\end{itemize}

\begin{itemize}
\item Scenarios IV-a: {M1A} + {sA} and V-a: {sA}. \\ We explore two distinct possibilities for the origin of stellar stream {sA}: first, that its progenitor, {M1A}, is still a visible component (Scenario IV-a), and second, that the progenitor was completely tidally disrupted, leaving only {sA} as a remnant (Scenario V-a) \citep{Deason2023UnravellingFunction, Panithanpaisal2021}. Figure \ref{fig:SceIVaSceVa} illustrates the best-fit orbits for both cases. {Scenario IV-a: {M1A} as Progenitor (Fig. \ref{fig:SceIVaSceVa}, up panels).} The best-fit orbit, integrated over $[-1.4, 1.7]\ \text{Gyr}$ and characterised by low eccentricity ($\epsilon \sim 0.08$), successfully connects {M1A} with the observed leading and trailing segments of {sA}. {Key Finding:} This represents a tight polar solution for the {M1A} progenitor, with the first passage occurring ${1.4\ \text{Gyr}}$ ago at a pericentre of approximately ${P \sim 15\ \text{kpc}}$, resulting in a maximum line-of-sight velocity variation of ${\Delta V_{\text{los}} \approx 400\ \text{km/s}}$ along the orbit. This scenario establishes a kinematic link between {M1A} and {sA}. {Scenario V-a: Disrupted Progenitor (Fig. \ref{fig:SceIVaSceVa}, bottom panels).} This hypothesis models {sA} as the remnant of a completely destroyed progenitor. The best-fit orbit, integrated on $[-1.3, 0.7]\ \text{Gyr}$, requires the progenitor to have undergone an extremely close pericentre passage, occurring ${1.2\ \text{Gyr}}$ ago at ${P \sim 3\ \text{kpc}}$ with a high velocity of ${|V_{\text{p}}| \approx 493\ \text{km/s}}$. {Key Finding:} The extremely small pericentre strongly supports the possibility of complete tidal disruption. Intriguingly, the proximity of the solutions (similar time frames and dynamics) suggests that the orbital characteristics of {sA} are strongly constrained by the streams regardless of the progenitor's current status (intact {M1A} \cRB{versus} destroyed progenitor).
\end{itemize}

\begin{itemize}
\item Scenario IV-b: {M1A} + {sA} + {sB}. \\ This situation posits {M1A} as the single progenitor for both stream {sB} and stream {sA}. {Key Finding:} It was not possible to find a single fitting orbit to connect both streams simultaneously. Moreover, the best solution found was an apparently {unbound} orbit, which is shown in panel iii), Fig. \ref{fig:notFits}. This high-energy orbit had a pericentre passage ${0.1\ \text{Gyr}}$ ago at ${24\ \text{kpc}}$ with a velocity of ${\sim 814\ \text{km/s}}$.
\end{itemize}

\begin{itemize}
\item Scenario V-b: {sB} (Disrupted Progenitor). \\ We investigate the possibility that {sB}'s progenitor has been completely destroyed by the tidal forces of Malin~1 \citep{Deason2023UnravellingFunction,Panithanpaisal2021}. The results are shown in panel v), Fig. \ref{fig:notFits}. The required passage to generate {sB} has a eccentric $\epsilon \sim 0.60$, occurring ${33\ \text{Myr}}$ ago, with a pericentre of ${P \sim 11\ \text{kpc}}$ and a high velocity of $D \sim 916\ \text{km/s}$. {Key Finding:} This small pericentre strongly indicates that complete tidal disruption is plausible. Intriguingly, the current position and trajectory of the modelled stream {sB} in this scenario point dynamically towards the position of the other progenitor candidate, {eM1}.
\end{itemize}

For a complete visualisation of the unbound solutions, see {Appendix \ref{app:disfavored}}. 

\subsection{{Evaluating orbital solutions}}
\label{sec:comparisons}
We illustrate three representative posteriors in the main text, chosen to span both halo \cRB{parametrizations} (ISO and NFW) and two orbital families (polar and radial). The strongest evidence for the origin of {Malin~1} streams comes from models dominated by the {ISO} (cored) \cRB{DM} halo, which features slightly more frequently among the top-ranked posterior probabilities (Table \ref{tab:DMcomparison}). To illustrate the spectrum of successful fits, we highlight three representative solutions.

\begin{table}[ht]
\centering
\caption{Best-fit posterior probabilities ($p$) for all mode\cRB{l}led scenarios.}
\label{tab:DMcomparison}
\tiny
\begin{tabular*}{\columnwidth}{@{\extracolsep{\fill}}lccccc@{}}
\hline\hline
Scenario & Orbit & B/U & DM Model & $\log(p)$ & $p$ \\
\hline
I-a   & Polar  & B & ISO & $-0.0934$ & $0.911$ \\
IV-a  & Polar  & B & ISO & $-0.1057$ & $0.900$ \\
V-a   & Radial & B & NFW & $-0.1056$ & $0.900$ \\
V-b   & Radial & U & ISO & $-0.1166$ & $0.888$ \\
V-a   & Radial & B & ISO & $-0.1216$ & $0.886$ \\
I-a   & Radial & B & ISO & $-0.1226$ & $0.885$ \\
III-a & Radial & U & NFW & $-0.1319$ & $0.877$ \\
IV-a  & Polar  & B & NFW & $-0.1528$ & $0.858$ \\
III-a & Radial & U & ISO & $-0.1564$ & $0.855$ \\
I-a   & Polar  & B & NFW & $-0.1769$ & $0.838$ \\
II-a  & Polar  & B & ISO & $-0.2471$ & $0.781$ \\
I-a   & Radial & B & NFW & $-0.2895$ & $0.749$ \\
II-b  & Radial & B & ISO & $-3.0291$ & $0.048$ \\
II-b  & Radial & B & NFW & $-4.4240$ & $0.012$ \\
\hline
\end{tabular*}
\tablefoot{
The probabilities are categorized by orbital configuration and Dark Matter (DM) halo model. Solutions are sorted by decreasing posterior probability. The column B/U denotes the gravitational status of the best-fit orbit: B for Bound and U for Unbound.
}
\end{table}

\begin{enumerate}
\item {Scenario I-a, polar, ISO:} This case provides the {statistically strongest evidence} among all solutions ($\log p = -0.0934$, $p \simeq 0.911$) and produces a clean posterior distribution with a narrow position angle (PA). {Key Finding:} This demonstrates the high viability of {polar orbits} when modelling the interaction of ${eM1}$ and ${sB}$ within a {cored halo} potential.

\item {Scenario IV-a, polar, ISO:} With $\log p = -0.1057$ ($p \simeq 0.900$), this solution is slightly less statistically favoured but achieves a more radial plunge. {Key Findings:} {In a cored halo, this radial increase offsets the weaker inner density slope, yet still yields enough tidal stripping at pericentre to generate Stream A} (${sA}$).

\item{Scenario V-a, radial, NFW:} Our baseline disruption solution, statistically related to Scenario IV-a ($\log p = -0.1056$, $p \simeq 0.900$), has a small pericentre ($R_p \simeq 9$ kpc) and $V_p \simeq 437$ km\,s$^{-1}$. {Key Finding:} In a {cuspy (NFW) halo}, such close high-velocity conditions strongly \cRB{favour} the necessary tidal stripping to explain the complete disruption of the progenitor and are highly consistent with the observed Stream A (${sA}$).
\end{enumerate}

We conducted a thorough orbital analysis that spans a wide ${40}$ model combination, covering the full spectrum of physically plausible configurations for \cRB{DM} halos (cusp versus core) and orbital geometries (polar versus radial). Although Table \ref{tab:DMcomparison} catalogues only the best-fit posterior probabilities ($p$) discovered during our search, the solutions presented encompass the most representative and dynamically relevant scenarios. We focus on this subset of solutions, as they successfully reproduce the observed stream track and line-of-sight velocity gradient, including complex cases where a single progenitor is fitted to multiple data sets. This is the Scenario II-b case, in which we used two different sets of data that could be the cause of a small posterior probability (see Table \ref{tab:DMcomparison}).

Using both NFW and ISO DM models, we found that bound orbital solutions were more likely under the ISO profile. This is supported by our maximum likelihood analysis, where the ISO model yielded systematically higher log-likelihood values in both the radial and polar scenarios, see Table~\ref{tab:DMcomparison}. This outcome likely arises because the ISO profile predicts a more extended mass distribution at large radii (Fig. \ref{fig:emceeFittings}, bottom panel), which better accommodates the gravitational constraints required by the satellite orbits. 
These findings are consistent with the results of \citep{Lelli2010}, who also uses ISO profiles to fit rotation curves, despite the general similarities with NFW-based models. 

In Bayesian terms, the values $\log(p)$ reported in Table~\ref{tab:DMcomparison} represent the logarithm of the posterior probability, combining the likelihood of the model with the prior distributions. Since we maintained consistent prior choices across both the ISO (cored) and NFW (cusped) \cRB{DM} halo models, the systematic trend toward\cRB{s} higher $\log(p)$ values for the ISO halo solutions reflects a quantitatively better fit to the available kinematic and spatial data. This evidence strongly suggests that a cored DM profile is more compatible with the observed orbital solutions of the Malin~1 streams.

It is important to note that the posterior probabilities in Table \ref{tab:DMcomparison} depend on the specific data sets used to constrain the models. For example, scenarios in which the progenitor must simultaneously satisfy the constraints of two distinct data sets, such as {Scenario II-b} (${M1C}+{sA}+{sB}$), incur a significant penalty in their \cRB{marginalised} likelihood, which is reflected in a lower overall $\log(p)$. Despite this statistical demerit, such solutions are of high physical significance, as they demonstrate the orbital ability of a single progenitor to reproduce multiple stellar features.

\section{Discussion}
\label{sec:discuss}

\subsection{Satellite galaxies and stellar streams.}
\label{sec:streams}
{Studies have shown a possible connection between the tidal formation and evolution of \cRB{compact elliptical} satellites and their host spiral galaxy, such as M32 and the Andromeda galaxy stream \citep{Ibata2004, Fardal2006}. 
This supports the hypothesis that the \cRB{compact elliptical} satellites M1A, M1B, and possibly M1C, could also be identified as progenitors of streams A and B as they are tidally processed.}
{Similar findings have been reported for other galaxies, for instance the giant galactic tail \textit{Kite}, which extends over more than $350$ kpc \citep{Zaritsky2023AnTail}, and the previously mentioned $\sim 500$ kpc giant stream in the Coma cluster \citep{Roman2023}. Both features support the presence of such large-scale streams.}

{In our study, we identify, among others, a possible orbital solution that connects almost linearly in projection the stream sB with eM1 and Malin~1. 
Moreover, the central region of eM1 appears off-centre (Fig. \ref{Fig:Malin1EnvironmentIn} - point 4), with an ongoing star formation, which would be naturally explained if it had a fly-by through Malin~1. Such phenomena are often observed in galaxies that undergo gravitational interactions, resulting in an off-centre core or a lopsided disc \citep{Mao2021}.} 
{Moreover, we are able to identify radial and polar orbital configurations for eM1. The radial configuration (panel i, Fig.~\ref{fig:SceIarSceIp}) had a $\sim33$~kpc pericentre passage$\sim 1.6$~Gyr ago, being now moving away from Malin~1.
More interestingly, the apocentre of this orbit happened 13 Gyr ago, 
implying that this satellite could be on its first infall.
We also estimated that the Jacobi tidal radius of eM1 at the pericentre was between $\sim 9-12$~kpc depending on the stellar and \cRB{DM} range. 
This would again naturally explain the observed internal disturbed morphology. 
Conversely, the polar configuration reaches a more distant pericentre of $150$~kpc, resulting in larger tidal range radii ($\sim 30-70$~kpc) and suggesting only mild tidal stripping. 
Furthermore, while polar configurations have been observed in systems \cRB{such us} the Sombrero galaxy \citep{Martinez-Delgado2021}, the radial interaction is physically analogous to the dramatic tidal stripping seen in the M31-M32 system \citep{Fardal2006}. 
Consequently, the radial orbit represents a more efficient mechanism for fueling the observed stellar streams and establishing eM1 as the primary progenitor of the sB stellar component.}

{We also identify another polar-like orbital solution perpendicular to Malin~1 dis\cRB{c}, where M1A could be connected to sA (scenario IV-a, panel a) in Fig.~\ref{fig:SceIVaSceVa}).
In this arrangement, the orbital solution is also radial having an apocentre of $\sim$100 kpc and a pericentre of $\sim 16.1 \pm 1.2$ kpc that could have impacted Malin~1 dis\cRB{c}s, providing a possible trigger mechanism for the active star formation observed in Malin~1 \citep{Lelli2010,Junais2024MUSEAttenuation,Junais2025}, and in the central region \citep{Johnston2024,Kataria2025}. 
We note that here we assumed that sA extends along its orbit. 
Another possibility is that sA could be a tidal arm structure stretching from M1A and aligned towards Malin~1 centre following the Lagrange points. Given that our simplistic orbital analysis cannot reproduce these tidal structures, we leave this for a follow-up N-body study.
Furthermore, our attempts to connect the satellite M1B with the streams sA and sB (Scenarios III-a/-b) resulted in unbound solutions due to the satellite's LoS velocity and position near Malin~1 centre, which forces it to have large tangential velocities to reach the streams orbital paths.}

{The satellite M1C appears to be a ${\rm H\alpha}$ blob \citep[][\footnote{private communication}]{Junais2024MUSEAttenuation}. 
In \citep{Ji2021} they propose that blobs in general could be tidal remnants or even progenitors of \cRB{UDGs}, which could be supported by M1C's observed high ${\rm H\alpha}$ velocity dispersion ($\sim90\kms$). 
We found orbits that could connect M1C to the streams, including an interesting case that connects both streams simultaneously (panel (a) of Fig. \ref{fig:SceIIb}). 
This solution offers a compelling solution already seen in other galactic systems, and would imply that the complete stream (panel (b) of Fig. \ref{fig:SceIIb}) could be even longer, extending 500 kpc long at least.}

{We explored orbital scenarios where we assume that the stellar streams are the disrupted progenitors themselves, as seen in other systems \citep{Deason2023UnravellingFunction, Panithanpaisal2021}. 
In scenarios V-a and V-b we identify possible radial orbital solutions for the streams sA and sB, which could imply recent tidal disruptions.
However, solution V-b produces an unbound orbit.
We note that in the solution V-a the stream sA appears disconnected from sB, suggesting that it is an independent event. Moreover, the spatial proximity of M1A, sA and M1B also suggests a complex interaction between Malin~1 and this \cRB{sub-group} of satellites.}

{Finally, we also identified unbound orbital solutions. 
For example, we attempted to connect eM1 with both stellar streams simultaneously, finding only unbound solutions, as shown by the best-fit trajectories for this configuration (Fig.~\ref{fig:notFits}, panel i). 
Similar unbound trajectories arise across multiple configurations, including (M1B, sA, sB), (M1A, sA, sB), (M1B, sB), and the sB-only case, as illustrated in panels ii) through v) of Fig.~\ref{fig:notFits}. 
These unbound solutions could suggest that the mass distribution of Malin~1 is inconsistent with the kinematic states of the satellites, implying that Malin~1 could be more massive than we can estimate by the rotation curve alone.
Consequently, given that $\Lambda\textsc{cdm}$ galaxy formation models show that galaxies are unlikely to have such orbits \citep[e.g.][]{Knebe2011}, we exclude these specific orbital configurations as viable evolutionary paths, implying that the observed stellar streams are likely the result of more constrained, bound accretion events rather than fast unbound encounters.}

\subsection{Satellite interactions and Malin~1 gLSB evolution.}
\label{sec:satint}

{Recent high-resolution observational studies of Malin~1 have revealed a complex dynamics and star formation history \citep{Galaz2015,Galaz2022, Galaz2024, Johnston2024,Junais2025,Kataria2025}, suggesting an active past involving mergers or accretion.}
{Studies have shown in general that examining the formation of tidal streams could provide insight into the timing of major merger events in the main host galaxy \citep[e.g.][]{Martinez-Delgado2023}.}
{\citet{Galaz2024} reported evidence of ongoing star formation in regions of extremely low molecular gas density, while \citet{Johnston2024} identified a possible double nucleus, pointing towards a recent disturbance or merger. 
This central activity is further supported by the H$\alpha$ emission studies by \citet{Junais2024MUSEAttenuation} 
and by the presence of high non-circular motions \citep{Kataria2025}.
Within this framework, our semi-analytical orbital analysis explores a wide set of configurations involving Malin~1, the two stream candidates sA and sB, and at least four satellite galaxies —M1A, M1B, M1C, and eM1—and consistently recovers kinematic signatures that corroborate these high velocity encounters.}

{These non-circular motions are often attributed to recent events such as stellar feedback and satellite accretion \citep{Saburova2021b}. Our derived orbital velocities (e.g., $V_{tg}=100-700$ km/s for many scenarios, see Table \ref{tab:scenarioranges}) are high enough to naturally explain the observed non-circular motions in the inner H$\alpha$ and [OII] components discussed by \citet{Kataria2025}, suggesting that the system's current high-velocity dispersion could be a direct consequence of these high-speed encounters considered in our models.}

{The observed state of Malin~1 likely reflects a multi-stage assembly process consistent with recent cosmological simulations. For example, \citet{Zhu2018} found a Malin~1 simulated analogue that formed of large diffuse galactic disc driven by the stimulated hot halo gas accretion  
and the accretion of a pair of intruding gas-rich galaxies in a coherent plane, as found in other simulated gLSBs \citep{Zhu2023}. 
Our orbital analysis offers a compelling candidate for this process: the temporal coincidence of the eM1 radial passage ($\sim -1.6$~Gyr) and the M1A encounter ($\sim -1.4$~Gyr) suggests that these satellites may have acted as intruding galaxies as found in \citet{Zhu2018}.
This sequential interaction likely catalyzed the dynamical disturbance and subsequent gas inflow necessary to form the observed stellar streams and support the expansive growth of the disc. Consequently, we propose that the morphological state of Malin~1 is the observational legacy of this merger-driven accretion event, which occurred between approximately $1.4$ and $1.6$~Gyr ago.}

{The morphological inspection of Malin~1 LSB dis\cRB{c} reveals a rich complexity, where we can specifically identify an optical cavity (Fig.~\ref{Fig:Malin1EnvironmentIn}, Point 7).
Additionally, \HI{} observations reveal a warp in the extended gaseous dis\cRB{c}, as well as a substructure near the cavity \citep{Pickering1997, Lelli2010}.
Simulations reveal that these types of features can be produced by gravitational interactions with satellite galaxies \citep[e.g.][]{Block2006,Gordon2006,Kalberla2009,Kim2014a}. Our orbital models indicate that these features are kinematically consistent with recent high-velocity encounters. Our orbital models for M1C reveal two distinct dynamical pathways that explain these structures: a polar configuration (Scenario II-a) with a large pericentre ($\sim 145$ kpc), which aligns with empirical warp models \citep{Binney2024}, and a radial disc-crossing configuration (Scenario II-b) with a tight pericentre ($\sim 3$ kpc) approximately $0.2$ Gyr ago. This recent radial passage provides the gravitational tide strength necessary to drive the formation of the observed stellar streams (sA and sB) and the central disc cavity, positioning M1C as a primary progenitor for these features.}
{Furthermore, the radial encounter of eM1, which occurred $\sim 1.6$~Gyr ago with a pericentre of $\sim33$~kpc, provides a compelling explanation for the larger-scale \HI{} warp and the current kinematic state of the outer disc.
Given that the rotation periods in the outer dis\cRB{c} are very slow, the cavity at $R\sim 50$ kpc is around $T_{\rm \theta}=2\pi R/ v_{\rm c}\sim1.6$~Gyr.
Therefore, the cavity could have already rotated around Malin~1 since the eM1 passage. 
However, independent of the impactor, we identify the cavity as a potential observational fossil substructure produced in an impulsive gravitational encounter.} 

\section{Conclusions and future work}
\label{sec:conclu}

{In this study, we have performed a comprehensive analysis of the stellar streams and satellite surrounding the \cRB{gLSBG} Malin~1. 
For this we  combined deep $R$-band photometry with HI and H$\alpha$ kinematics to determine the gravitational potential of Malin~1 to model the orbital dynamics of its satellites \cRB{and} search for orbital solutions that can connect them with the observed stellar streams sA and sB \citep{Galaz2015}. Our key findings are summari\cRB{s}ed as follows:}
\begin{itemize}
    \item[-] {Massive DM halo:} {We confirm that Malin~1 is hosted by a massive \cRB{DM} halo (Section \ref{sec:res:potmod}). The MCMC model\cRB{l}ing of the rotation curve yield\cRB{ed} a virial mass of $M_{\text{vir}} = 1.4^{+0.3}_{-0.2} \times 10^{12} M_{\odot}$ for an NFW profile and $M_{\text{vir}} = 2.6 \pm 0.4 \times 10^{12} M_{\odot}$ for a pseudo-Isothermal profile. The latter provides a statistically better fit to the outer rotation curve data.}
    \item[-] {Connecting satellite orbits with streams:} {\cRB{U}sing our orbital models, we identified viable orbital solutions\cRB{, which} we call scenarios (Section \ref{sec:res:scen}) connecting the satellite galaxy candidates  M1A, M1B, M1C, and eM1 to the observed stellar streams. These \cRB{scenarios are ase follows}:}

\begin{itemize}
    \item[-] {Malin~1 interacting with satellite M1A provides a bound orbital configuration, identifying M1A as a plausible progenitor \cRB{of} the stellar stream sA.}
    
    \item[-] {Interaction models involving \cRB{the} satellite M1B suggest that it is a candidate progenitor for the streams; however, the resulting orbits are unbound or marginally bound implying high-velocity encounters.}
    
    \item[-] {The orbital connection between \cRB{the} satellite M1C and stream sB can be modelled via a polar configuration. We also identif\cRB{ied} a radial orbital solution that allows streams sA and sB \cRB{to simultaneously connenct} with M1C (Fig.\ref{fig:SceIIb}), which would imply that the stream is more extended than currently observed.}

    \item[-] {We identif\cRB{ied} two types of orbits that can connect the distant eM1 satellite with sB, finding radial and polar configurations (Fig.~\ref{fig:SceIarSceIp}). 
    Both solutions result in long orbital periods, with the radial solution allowing stronger tidal interaction with Malin~1 and indicating that eM1 would be on its first infall.}
    
\end{itemize}

{These results indicate that while several satellites could be progenitor candidates of the observed streams and tidal debris, their specific interaction properties (e.g. pericentres, bound \cRB{versus} unbound) vary significantly across the system.}

\item[-] {Types of orbital interactions:} {Our \cRB{modelling revealed} two distinct modes of satellite interaction categori\cRB{s}ed by orbital geometry and pericentric proximity:}

{
\begin{itemize}
\item {Close radial encounters:} These represent high-impact trajectories characterised by small pericentres. Notable examples include the M1C-driven interaction (driving sA and sB) and the sA-only event both occurring at $\sim 3$~kpc as found by \citet{Johnston2024,Kataria2025}, as well as the M1A-driven interaction at $\sim 15$~kpc and the eM1 radial encounter at $\sim 33$~kpc. These close-range crossings generate the impulsive gravitational forcing necessary to trigger significant tidal stripping and drive the observed morphological evolution of the Malin~1 disc.
\item {Distant polar orbits:} In contrast, we also identif\cRB{ied} possible polar orbital configurations, similar to the ones observed in the Sombrero galaxy \citep{Martinez-Delgado2021}. These solutions involve the satellites eM1 \cRB{and} M1C connected with the stream sB, and \cRB{they} exhibit larger pericentres of $\sim 145$--$147$~kpc. These solutions imply weaker interactions with \cRB{the outer halo of} Malin~1 rather than direct crossing events with the HSB\cRB{,} LSB, \cRB{or both}  disc components.
\end{itemize}
}
\end{itemize}

{The findings of this work provide a viable dynamical framework for understanding the role of 'nurture' in the evolution of \cRB{gLSBGs}. To further refine this history, future work will focus on \cRB{the areas discussed below}:}

\begin{itemize}
\item[-] {Spectroscopic analysis:} {Upcoming MUSE and JWST data will be utili\cRB{s}ed to determine the metallicities and star formation histories of the streams and M1C. This will help distinguish between tidal debris and \textit{in situ} star-forming regions (e.g. H$\alpha$ blobs).}
\item[-] {Simulations:} {The orbital parameters derived in this study will serve as initial conditions for full N-body hydrodynamical simulations. These will investigate the long-term survival of the streams and the origin of the observed warp and cavity in the outer disc.}
\item[-] {Environmental survey:} {A broader analysis of the 1~Mpc environment around Malin~1 will clarify the impact of the local cosmic web on the galaxy's continued growth and satellite accretion history (Bustos-Espinoza et al. in prep.).}
\end{itemize}

{These dynamical interactions likewise offer a solid basis for interpreting the recent star formation history of Malin~1. The impulsive gravitational encounters found in our models—especially those characteri\cRB{s}ed by small-pericentre radial passages—are highly effective at driving gas inflows and concentrated star formation, in agreement with recent observational work \citep{Galaz2022, Galaz2024,Johnston2024,Kataria2025}. This implies that the observed central activity and irregular star-forming patches are not random \cRB{but that they} instead arise directly from satellite-driven gas compression and tidal perturbations.}

\cRB{Our dynamical modeling successfully constrains the diverse orbital configurations capable of producing the observed stellar features.} Ultimately, this study shows that the environment of Malin~1 is complex, and it could \cRB{be the} result of multiple satellite tidal interactions, providing a key benchmark for the role of hierarchical assembly in the evolution of \cRB{gLSBGs}.

\begin{acknowledgements}
This paper is part of a Ph.D. thesis. The primary author expresses gratitude for support received from the ANID BASAL project CATA FB210003 and Pontificia Universidad Católica de Chile (PUC), as well as access to the Horus and Geryon 2 PUC servers. Support was also received from the Organization of American States (OAS) for the 2021-2022 scholarship and from the Sociedad Chilena de Astronomía (SOCHIAS) through two grants under the Prof. Adelina Gutiérrez program. Furthermore, the author acknowledges support from the Carrera de Física at the Instituto de Investigaciones Físicas, Universidad Mayor de San Andrés (UMSA) in La Paz, Bolivia, and is grateful to the Indian Institute of Astrophysics for a six-week visit, including the Vainu-Bappu Observatory, in early 2025. {This research was also supported by the Asociación Boliviana para el Avance de la Ciencia (ABAC) and Sociedad Boliviana de Física (SOBOFI)}. MD and SB acknowledge the support of the Science and Engineering Research Board (SERB) Core Research Grant CRG/2022/004531 and the Department of Science and Technology (DST) grant DST/WIDUSHI-A/PM/2023/25(G) for this research. The authors (RBE, GG, and MB) acknowledge the support of the European Southern Observatory (ESO) for providing space for discussions while this paper was written. GG thanks Fondecyt Regular 1230231. GG thanks Pontificia Universidad Católica de Chile, ESO, France-Chile Laboratory of Astronomy (FCLA), and Laboratoire d'Astrophysique de Marseille (LAM) for their support during a 2024-2025 sabbatical leave. J. is funded by the European Union (MSCA EDUCADO, GA 101119830 and WIDERA ExGal-Twin, GA 101158446). THP acknowledges support from the Agencia Nacional de Investigación y Desarrollo (ANID) grant CATA-Basal FB210003. This research has used the NASA/IPAC Extragalactic Database (NED), which is operated by the Jet Propulsion Laboratory, California Institute of Technology, under contract with the National Aeronautics and Space Administration. This research has made use of NASA's Astrophysics Data System Bibliographic Services.
\end{acknowledgements}


\begin{appendix}

\section{Stream points}
\label{app:streampoints}

Table \ref{Table:streamPoints} presents the stream points used as prior information for the Monte Carlo Markov chain \textsc{mcmc-emcee} method \citep{Foreman2012}, identified visually from Fig.~\ref{Fig:Malin1EnvironmentIn} \citep{Galaz2015}.

\begin{figure}[ht]
   \centering
   \includegraphics[width=\linewidth]{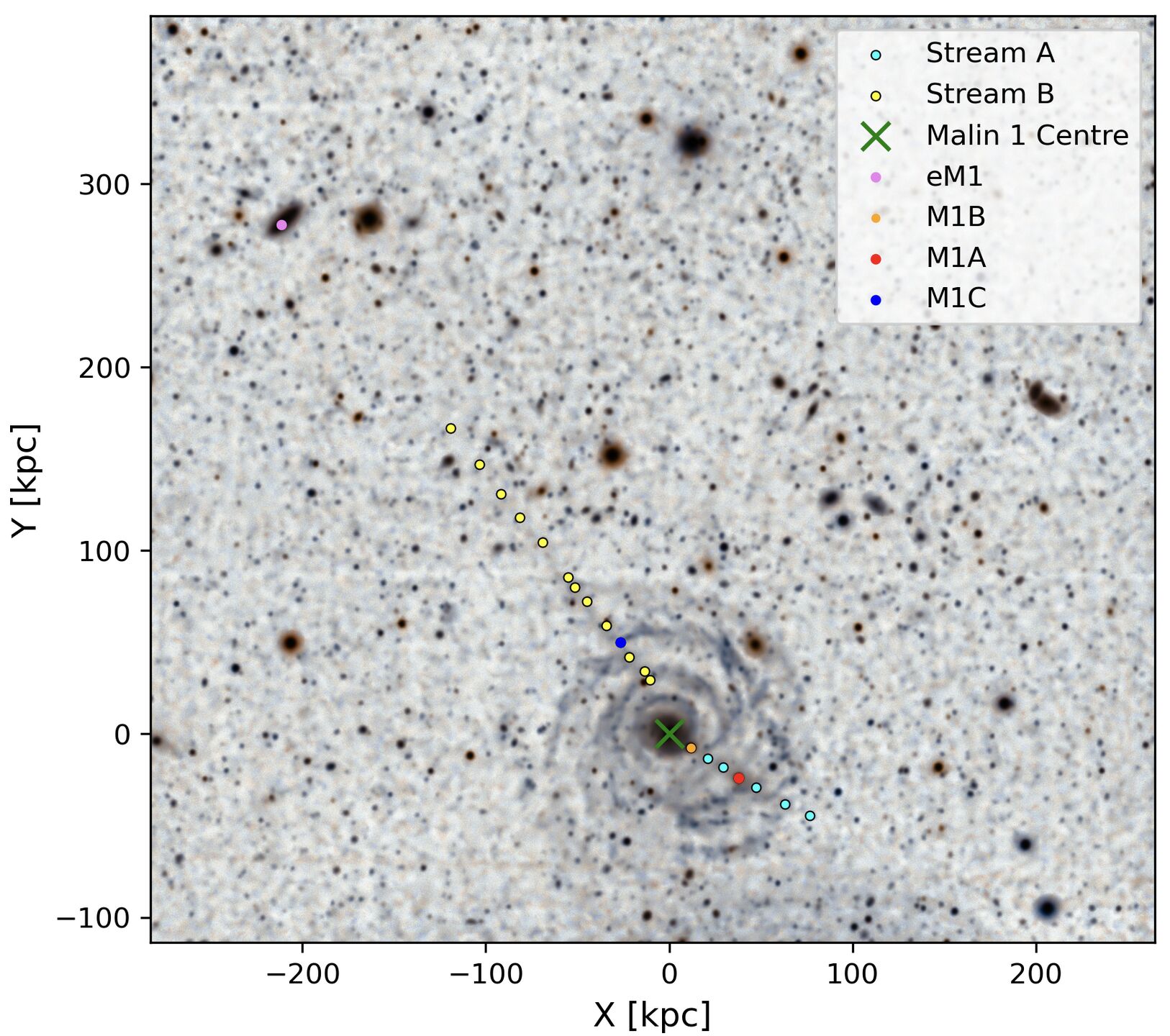}
    \caption{Stream points sA (cyan points) and sB (yellow points), and four satellites, M1A (red), M1B (orange), M1C (blue), and eM1 (violet). The centre of Malin~1 is marked with a green cross. Data values are shown in Table \ref{Table:streamPoints}.}
    \label{Malin1_AandBstreamPoints}
\end{figure}

\begin{figure}[ht]
   \centering
   \includegraphics[width=\linewidth]{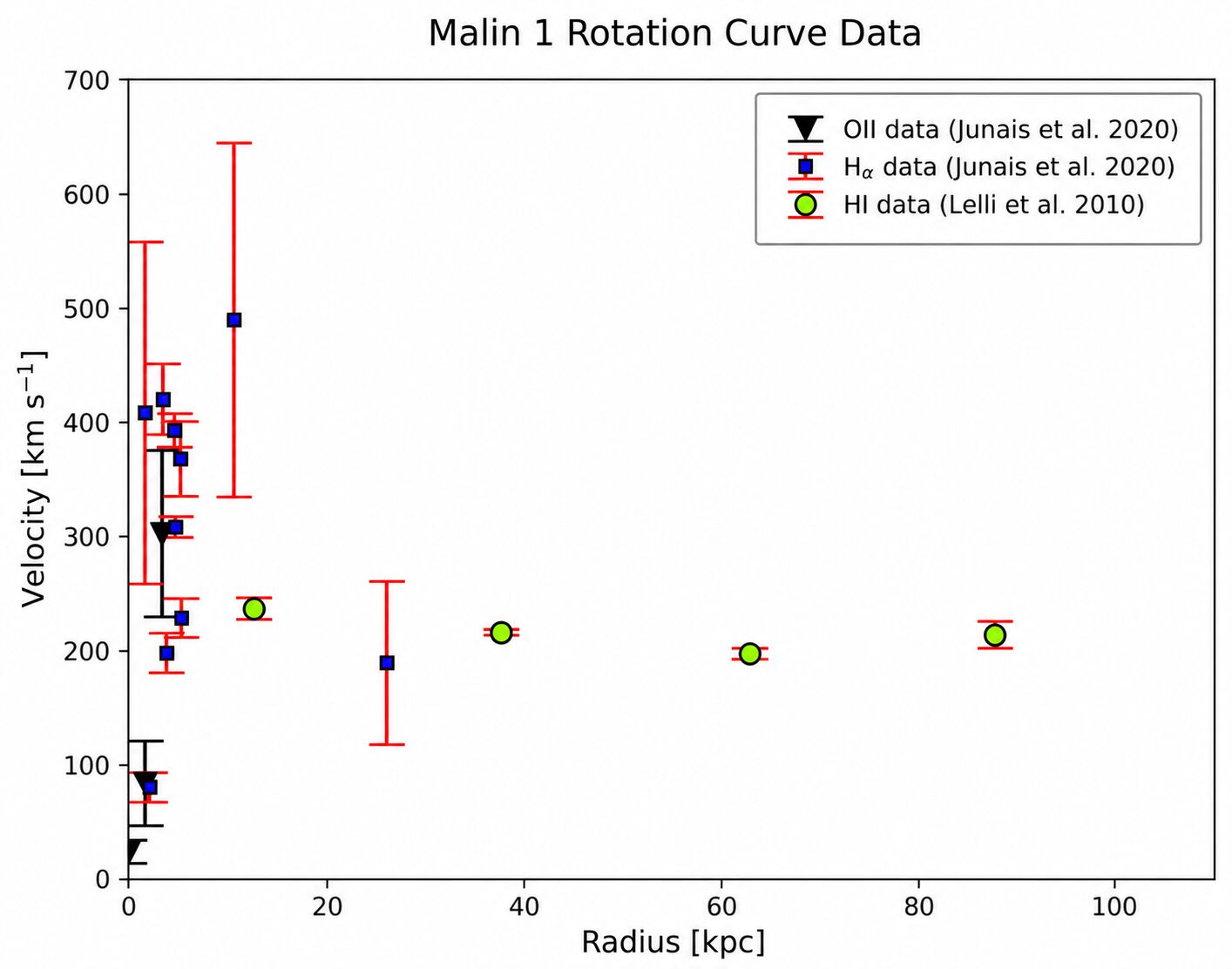}
    \caption{Observed Malin~1 rotation curve including \ion{H}{i}, OII, and H$\alpha$ data.}
    \label{Malin1_RCwHIHOIIHalpha}
\end{figure}

\begin{table*}[ht!]
\centering
\caption{Stream points sA and sB coordinates.}
\label{Table:streamPoints}
\tiny
\begin{tabular*}{\textwidth}{@{\extracolsep{\fill}}lccccc@{}}
\hline\hline
Object & $\alpha$ (J2000) & $\delta$ (J2000) & $x$ [kpc] & $y$ [kpc] & $r$ [kpc] \\
\hline
Malin~1 & 12:36:59.36 & 14:19:49.4 & $0 \pm 1$ & $0 \pm 1$ & $0$ \\
\noalign{\smallskip}
sA1: M1B & 12:36:58.89 & 14:19:43.9 & $12 \pm 1$ & $-8 \pm 1$ & $14 \pm 1$ \\
sA2 & 12:36:58.42 & 14:19:39.4 & $21 \pm 1$ & $-13 \pm 1$ & $25 \pm 1$ \\
sA3 & 12:36:58.03 & 14:19:36.1 & $30 \pm 1$ & $-18 \pm 1$ & $35 \pm 1$ \\
sA4: M1A & 12:36:57.62 & 14:19:32.9 & $38 \pm 1$ & $-24 \pm 1$ & $45 \pm 1$ \\
sA5 & 12:36:57.12 & 14:19:28.2 & $47 \pm 1$ & $-29 \pm 1$ & $56 \pm 1$ \\
sA6 & 12:36:56.78 & 14:19:26.8 & $63 \pm 1$ & $-38 \pm 1$ & $74 \pm 1$ \\
sA7 & 12:36:56.47 & 14:19:24.9 & $77 \pm 1$ & $-44 \pm 1$ & $89 \pm 1$ \\
\noalign{\smallskip}
sB1 & 12:36:59.98 & 14:20:07.1 & $-11 \pm 1$ & $29 \pm 1$ & $31 \pm 1$ \\
sB2 & 12:37:00.02 & 14:20:12.1 & $-13 \pm 1$ & $34 \pm 1$ & $37 \pm 1$ \\
sB3: M1C & 12:37:00.36 & 14:20:16.8 & $-22 \pm 1$ & $42 \pm 1$ & $47 \pm 1$ \\
sB4 & 12:37:00.65 & 14:20:22.6 & $-27 \pm 1$ & $50 \pm 1$ & $57 \pm 1$ \\
sB5 & 12:37:01.03 & 14:20:27.6 & $-34 \pm 1$ & $59 \pm 1$ & $69 \pm 1$ \\
sB6 & 12:37:01.34 & 14:20:34.4 & $-45 \pm 1$ & $72 \pm 1$ & $85 \pm 1$ \\
sB7 & 12:37:01.68 & 14:20:41.3 & $-52 \pm 1$ & $80 \pm 1$ & $95 \pm 1$ \\
sB8 & 12:37:01.70 & 14:20:41.6 & $-55 \pm 1$ & $85 \pm 1$ & $102 \pm 1$ \\
sB9 & 12:37:02.69 & 14:20:56.4 & $-69 \pm 1$ & $105 \pm 1$ & $125 \pm 1$ \\
sB10 & 12:37:03.22 & 14:21:02.9 & $-81 \pm 1$ & $118 \pm 1$ & $144 \pm 1$ \\
sB11 & 12:37:03.55 & 14:21:12.2 & $-92 \pm 1$ & $131 \pm 1$ & $160 \pm 1$ \\
sB12 & 12:37:04.10 & 14:21:24.8 & $-103 \pm 1$ & $147 \pm 1$ & $180 \pm 1$ \\
sB13 & 12:37:04.85 & 14:21:37.8 & $-119 \pm 1$ & $167 \pm 1$ & $205 \pm 1$ \\
\noalign{\smallskip}
eM1 & 12:37:08.92 & 14:22:53.3 & $-211 \pm 1$ & $278 \pm 1$ & $349 \pm 1$ \\
\hline
\end{tabular*}
\tablefoot{
Observational inputs for stream sA (7 points) and sB (13 points) used in the \textsc{mcmc-emcee} analysis \citep{Foreman2012}. Satellite positions are also included. Spatial locations of these points and companion satellites are illustrated in Fig.~\ref{Malin1_AandBstreamPoints}. Spatial uncertainties are set by the projected angular scale and median seeing of $0.8''$ \citep{Galaz2015}, corresponding to $\pm 1$ kpc. Radial distance errors are propagated.
}
\end{table*}

\section{Parameter priors and ranges}
\label{app:priors}

Table \ref{tab:scenarioranges} defines each orbital parameter: line-of-sight coordinate $z$, tangential velocity $V_{\mathrm{tg}}$, position angle PA, pericentre, apocentre, eccentricity $\epsilon$, and posterior log-probability. The astrophysical motivation for the parameter ranges is also explained.

\begin{table*}[ht!]
\centering
\caption{Best-fitted parameter ranges and astrophysical motivation for orbital scenarios.}
\label{tab:scenarioranges}
\tiny
\begin{tabular*}{\textwidth}{@{\extracolsep{\fill}}lccc@{}}
\hline\hline
Scenario & $z$ [kpc] & $V_{\mathrm{tg}}$ [km\,s$^{-1}$] & PA [deg] \\
\hline
I-a-Radial: eM1+sB & $-294.43_{-302.12}^{+172.90}$ & $250.27_{-130.54}^{+162.21}$ & $40.00_{-0.30}^{+1.48}$ \\
I-a-Polar: eM1+sB  & $1.72_{-30.18}^{+28.85}$     & $112.79_{-88.64}^{+221.07}$  & $41.00_{-1.41}^{+8.25}$ \\
II-b: M1C+sA+sB    & $0.10_{-2.52}^{+2.77}$       & $384.82_{-10.21}^{+7.21}$    & $37.00_{-0.54}^{+0.56}$ \\
III-a: M1B+sB      & $-0.32_{-7.80}^{+9.02}$      & $781.48_{-73.86}^{+77.08}$   & $27.00_{-3.97}^{+4.03}$ \\
IV-a: M1A+sA       & $10.85_{-13.96}^{+13.17}$    & $371.88_{-89.19}^{+86.88}$   & $240.00_{-2.66}^{+2.61}$ \\
V-a: sA            & $-0.15_{-6.87}^{+7.12}$      & $648.41_{-176.05}^{+171.08}$ & $-240.00_{-1.51}^{+1.56}$ \\
V-b: sB            & $-0.19_{-6.68}^{+6.93}$      & $703.99_{-138.87}^{+133.72}$ & $37.00_{-0.86}^{+0.82}$ \\
\hline
\end{tabular*}
\tablefoot{
Refer to Table~\ref{tab:DMcomparison} for the complete tabulated best-fit parameters and Table~\ref{tab:scenarioranges2} for the $R_{\mathrm{peri}}$, $R_{\mathrm{apo}}$, and $\epsilon$ ranges. 
{Astrophysical motivation:} 
{I-a-Radial:} Satellite projected near NE disc; PA limited to quadrant I based on stream direction; negative $z$ assumed. 
{I-a-Polar:} Polar orbit; stream as projection of circular orbit; orientation suggests infall from below. 
{II-b:} Highly unconstrained distant satellite; wide ranges used for exploration. 
{III-a:} Orbit near disc plane on southern side; PA reflects motion direction. 
{IV-a:} Stream features suggest decaying retrograde orbit entering from behind disc. 
{V-a:} Curvature favors transverse westward orbit; both $z$ signs allowed. 
{V-b:} Wide $z$ range tests vertical impact; PA constrained by stream bending.
}
\end{table*}

\begin{table*}[ht!]
\centering
\caption{Best-fit orbital geometry parameters and astrophysical motivations.}
\label{tab:scenarioranges2}
\tiny
\begin{tabular*}{\textwidth}{@{\extracolsep{\fill}}lccc@{}}
\hline\hline
Scenario & $R_{\mathrm{peri}}$ [kpc] & $R_{\mathrm{apo}}$ [kpc] & $\epsilon$ \\
\hline
I-a-Radial: eM1+sB & $182.05_{-72.17}^{+238.77}$ & $1736.05_{-1137.98}^{+1681.03}$ & $0.37_{-0.10}^{+0.19}$ \\
I-a-Polar: eM1+sB  & $159.90_{-17.44}^{+41.98}$  & $416.76_{-63.05}^{+1841.46}$  & $0.11_{-0.02}^{+0.34}$ \\
II-b: M1C+sA+sB    & $199.36_{-2.73}^{+2.51}$    & $1434.19_{-40.84}^{+28.70}$   & $0.35_{-0.01}^{+0.01}$ \\
III-a: M1B+sB      & $215.44_{-13.64}^{+17.32}$  & $682.37_{-40.34}^{+50.86}$    & $0.14_{-0.02}^{+0.03}$ \\
IV-a: M1A+sA       & $202.08_{-12.49}^{+13.04}$  & $565.20_{-60.28}^{+59.18}$    & $0.12_{-0.02}^{+0.02}$ \\
V-a: sA            & $211.30_{-7.67}^{+7.20}$    & $871.29_{-226.89}^{+286.35}$  & $0.21_{-0.07}^{+0.07}$ \\
V-b: sB            & $207.23_{-3.60}^{+3.69}$    & $944.86_{-225.36}^{+223.98}$  & $0.23_{-0.07}^{+0.05}$ \\
\hline
\end{tabular*}
\tablefoot{
Final best-fit orbital geometry parameters---pericentre ($R_{\mathrm{peri}}$), apocentre ($R_{\mathrm{apo}}$), and eccentricity ($\epsilon$)---for each scenario, corresponding to ISO and NFW dark matter profiles.
{Astrophysical motivation:} 
{I-a-Radial:} Satellite projected near NE disc; PA limited to quadrant I based on stream direction; negative $z$ assumed. High eccentricity ($\epsilon \approx 0.37$) and large $R_{\mathrm{apo}} \approx 1.7$\,Mpc indicate a loosely bound, highly elongated orbit. 
{I-a-Polar:} Polar orbit; stream as projection of circular orbit; orientation suggests infall from below. Nearly circular ($\epsilon \approx 0.11$) with the closest $R_{\mathrm{peri}} \approx 160$\,kpc. Apocentre uncertainty is the highest. 
{II-b:} Highly unconstrained distant satellite; wide ranges used for exploration. Highly elliptical ($\epsilon \approx 0.35$) with $R_{\mathrm{apo}} \approx 1.4$\,Mpc. Low uncertainty on $\epsilon$ suggests a robustly elongated shape. 
{III-a:} Orbit near disc plane on southern side; PA reflects motion direction. Compact, nearly circular ($\epsilon \approx 0.14$) orbit with moderate $R_{\mathrm{apo}} \approx 682$\,kpc, suggesting a tightly bound satellite. 
{IV-a:} Stream features suggest decaying retrograde orbit entering from behind disc. This is the most compact orbit overall, with the smallest $R_{\mathrm{apo}} \approx 565$\,kpc and very low eccentricity ($\epsilon \approx 0.12$). 
{V-a:} Curvature favors transverse westward orbit; both $z$ signs allowed. Moderately eccentric ($\epsilon \approx 0.21$). $R_{\mathrm{peri}}$ is well-constrained, but $R_{\mathrm{apo}}$ shows large uncertainties. 
{V-b:} Wide $z$ range tests vertical impact; PA constrained by stream bending. Moderately eccentric ($\epsilon \approx 0.23$) orbit. $R_{\mathrm{peri}}$ is one of the most precisely determined values in the table.
}
\end{table*}

\section{\textsc{mcmc} parameter fits}
\label{app:mcmc}

\subsection{NFW halo}

\begin{figure}[ht!]
   \centering
   \includegraphics[width=0.95\linewidth]{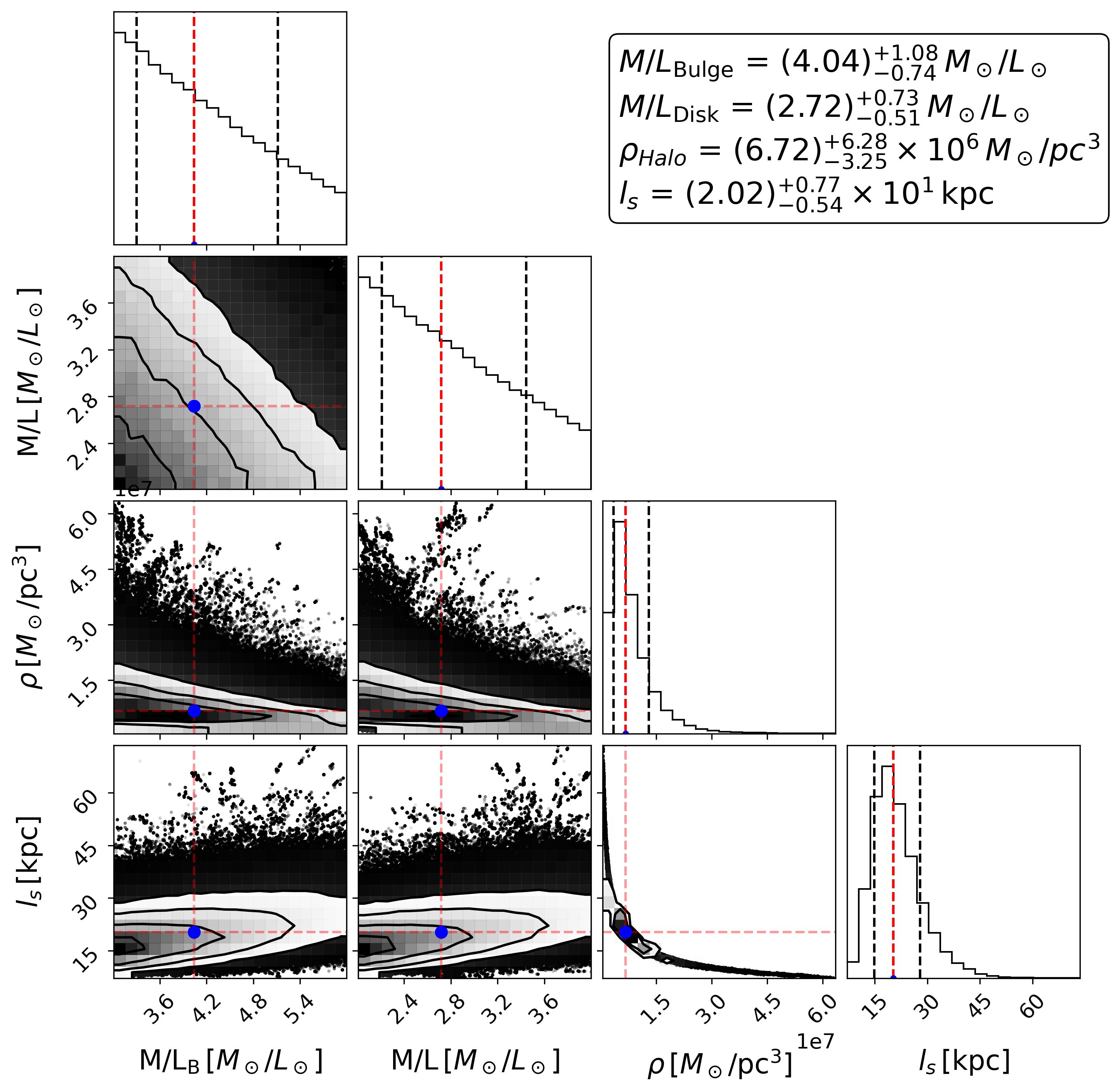}
   \caption{Posterior distributions of the mass-to-light ratios for the bulge and HSB disc, together with the NFW DM halo parameters (central density $\rho_{\mathrm{Halo}}$ and scale length $l_s$). Histograms (diagonal) show marginali\cRB{s}ed distributions with the 16th, 50th, and 84th percentiles. Contours display covariances. Best-fit values with $1\sigma$ uncertainties are reported in the legend. In the NFW case, maximum a posteriori (MAP) values deviate from the medians, highlighting weaker convergence compared to the ISO model in the main text.}
   \label{fig:ThreeMassToLightRatiosForNFW}
\end{figure}

\section{Unbound scenarios}
\label{app:disfavored}

\subsection{Unbound solutions}
Representative unbound or disfavo\cRB{u}red configurations are shown in Fig.~\ref{fig:notFits}. These include panel i) scenario I-b: eM1+sB+sA, ii) scenario III-b: M1B+sB+sA, iii) scenario IV-b: M1A+sA+sB, iv) scenario III-a: M1B+sB, and v) scenario V-b: progenitor destroyed; these could not satisfy the binding or tidal feasibility criteria. These fail because the total orbital energy is positive ($E > 0$) or because the tidal radius at pericentre exceeds the half-light radius of the progenitor, making tidal stripping inefficient. Such cases help illustrate the explored parameter space but are excluded from the main analysis.

\begin{figure*}[!ht]
   \centering
   \includegraphics[width=0.88\textwidth]{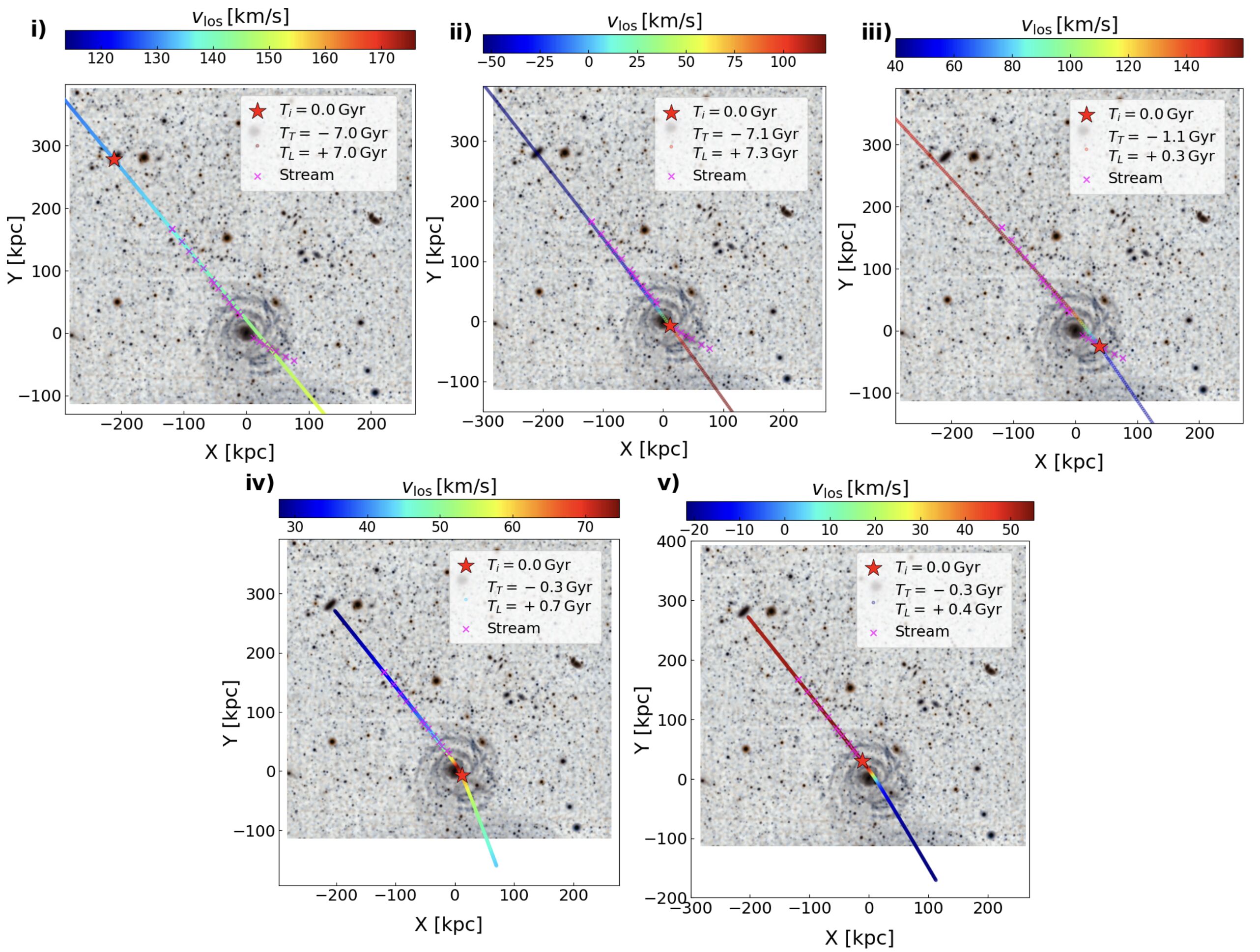}
   \caption{Examples of unbound or disfavo\cRB{u}red configurations. Panel i) scenario I-b: eM1+sA+sB, ii) scenario III-b: M1B+sA+sB, iii) scenario IV-b: M1A+sA+sB, iv) scenario III-a: M1B+sB, and v) scenario V-b: progenitor destroyed; these fail to satisfy binding or tidal feasibility criteria.}
   \label{fig:notFits}
\end{figure*}

\end{appendix}


\end{document}